\documentclass[twocolumn]{aa}

\usepackage{graphicx}

\usepackage{txfonts}
\usepackage[normalem]{ulem}

\usepackage{xcolor}
\usepackage[utf8]{inputenc}
\definecolor{xlinkcolor}{cmyk}{1,0.6,0,0}

\usepackage[breaklinks=true, 
    bookmarks=false,         
     pdfnewwindow=true,      
     colorlinks=true,    
     linkcolor=xlinkcolor,     
     citecolor=xlinkcolor,     
     filecolor=xlinkcolor,  
     urlcolor=xlinkcolor,      
final=true
]{hyperref}

\usepackage{subcaption}

\begin{document}

   \title{Puzzling Ultra-Diffuse Galaxy Evolution (PUDGE)}

   \subtitle{II. A transformation pathway from ultra-diffuse galaxies to compact dwarfs in galaxy clusters}

   \author{Nata{\v s}a Pavlov\inst{1}
          \and
          Ana Mitra{\v s}inovi{\' c}\inst{2,3}
          }

   \institute{$^1$ Faculty of Mathematics, University of Belgrade, Studentski trg 16, 11158 Belgrade, Serbia\\
              \email{natasa.pavlov@matf.bg.ac.rs}\\
              $^2$Astronomical Observatory, Volgina 7, 11060 Belgrade, Serbia\\
              \email{amitrasinovic@aob.rs}\\ 
              $^3$ Institute of Astrobiology, University Research and Innovation Center, University of Crete, Voutes Campus, 70013 Heraklion, Greece
               }

   \date{Received ; accepted }

  \abstract {Ultra-diffuse galaxies (UDGs) and compact dwarfs (CDs) occupy opposite extremes of the structural parameter space of dwarf galaxies, yet their spatial distributions in clusters suggest a possible evolutionary connection. Observational studies have reported a pronounced anticorrelation between the two populations interpreted as evidence that CDs represent tidally stripped remnants of diffuse progenitors, a scenario that implicitly assumes a compact stellar nucleus must be present at infall to survive environmental processing. We tested this hypothesis using the IllustrisTNG cosmological simulation (TNG100) by examining the UDG and CD populations in seven galaxy clusters and tracing the evolutionary histories of 112 present-day CDs. We confirm that TNG100 reproduces the observed spatial anticorrelation, with CDs concentrated within $d/R_{200} \lesssim 0.2$ and UDGs preferentially inhabiting the cluster outskirts. Tracing CDs back in time, we identified eight systems whose progenitors undergo a transient UDG phase, with extremely high gas fractions ($f_{\rm gas} \gtrsim 0.8$), immediately before cluster infall. In all eight systems, a vast majority of the present-day stellar mass was assembled after the epoch of maximum spatial extent, and the peak star formation rate during the transformation is the highest each galaxy achieves across its entire lifetime. The UDG progenitors show no prominent stellar cores before infall, demonstrating that the compact component of the resulting CD is not an exposed preexisting nucleus but is instead freshly built through starburst-driven star formation during the stripping process itself. Our results reveal a physically motivated UDG-to-CD transformation pathway driven entirely by cluster environment that is fundamentally distinct from classical tidal stripping scenarios and highlight the critical role of gas richness as a prerequisite for this channel. Accounting for roughly 7\% of our sample, the UDG progenitors and their evolution imply a possible route of CD formation.}


   \maketitle

\section{Introduction}

In the late or present-day Universe, matter is not spread randomly or uniformly but is instead organized into an intricate large-scale structure (LSS) called the cosmic web. This LSS consists of long, thread-like filaments of dark matter, gas, and galaxies, separated by enormous low-density regions called voids, which are sometimes bounded by flatter, sheet-like structures known as walls. The entire structure is interconnected, and the points where massive filaments intersect are called nodes. These nodes are the highest-density regions in the Universe, more commonly known as galaxy clusters \citep[see][for reviews of cluster formation and evolution]{Voit2005RvMP...77..207V, Kravtsov+Borgani2012ARA&A..50..353K}. Galaxy clusters are important and unique laboratories that enable us to study cosmology, galaxy evolution, and fundamental physics simultaneously. Thus, it is not surprising that modern cosmological simulations are moving toward focusing cluster modeling and analysis in greater detail, such as the MillenniumTNG simulations\footnote{\url{https://www.mtng-project.org/}} \citep[e.g.,][]{Bose+2023MNRAS.524.2579B, Pakmor+2023MNRAS.524.2539P} or the TNG-Cluster\footnote{\url{https://www.tng-project.org/cluster/}} \citep[e.g.,][]{Ayromlou+2024A&A...690A..20A, Lehle+2024A&A...687A.129L, Nelson+2024-tngcluster, Rohr+2024A&A...686A..86R}. 

When it comes to galaxy evolution, clusters are the most extreme environments in the Universe, and they have a profound, violent effect on any galaxy that resides in or falls into them. The standard cosmological model postulates that dark matter halos and their associated galaxies grow hierarchically, primarily through mergers with other galaxies. In high-density environments (such as clusters), galaxy mergers were much more frequent throughout the history of the Universe, shaping galaxy evolution and leading to the formation of red elliptical galaxies without star formation. Consecutive mergers of elliptical galaxies form massive elliptical galaxies, or present-day so-called central dominant galaxies, the most massive galaxies ever discovered \citep[e.g.,][]{Matthews+1964}, which are exclusively found in the centers of galaxy clusters. Additionally, densely populated galaxy clusters represent ideal environments for frequent nonmerger interactions, particularly close and penetrating encounters between satellite galaxies \citep{tormen1998, Knebe2004PASA}. The pioneering works in the study of galaxy interactions and their impact on observed morphology include \citet{tt1972} and \citet{Eneev1973}, who demonstrated that close encounters between galaxies lead to the formation of a variety of characteristic morphological structures, the ideas that were later confirmed by subsequent studies \citep[e.g.,][]{barnes&hernquist1992, Tutukov2006, Dubinski&Chakrabarty2009}. One particular type of galaxy interaction is flybys, very close nonmergers \citep{sinha2012}, which can strongly influence galaxy morphology \citep[e.g.,][]{Pettitt+Wadsley2018, Mitrasinovic+Micic2023}. These interactions can play an important role in the evolution of galaxies in rich clusters, where, due to the high relative velocities between galaxies, they typically significantly outnumber mergers in the late Universe \citep{shan2019}. However, galaxies in clusters also experience high-speed encounters at relatively large separations. This process, termed galaxy harassment \citep{moore1996}, though seemingly weak, can cause substantial morphological transformations, even driving galaxies from one morphological class to another \citep{moore1998, moore1999, mastropietro2005}.

Another important process in galaxy clusters is ram pressure stripping (RPS), which occurs when the interstellar medium of a galaxy interacts with the surrounding intracluster medium \citep{Boselli+2022A&ARv..30....3B}. This mechanism efficiently removes galactic gas, quenching star formation and potentially driving the transformation of spiral galaxies into lenticular (S0) ones. Gas removal and suppression of star formation due to RPS can occur in both group and cluster environments \citep[e.g.,][]{Reynolds2022, Lin2023, Holwerda2023, Holwerda2025}, and even at the cluster outskirts \citep[e.g.,][]{Lopes+2024MNRAS.527L..19L, Piraino-Cerda+2024MNRAS.528..919P}. There is also evidence of RPS efficiency depending on the stellar mass of a galaxy \citep{Xie+2025A&A...698A..73X}, with low-mass satellite galaxies being the most affected, while massive galaxies can still retain their gas and maintain star formation.

The extreme sensitivity of low-mass (satellite) galaxies to these environmental mechanisms results in a diverse population of dwarf galaxies within clusters. Among the most enigmatic members of this population are two classes that lie at opposite extremes of the structural parameter space: ultra-diffuse galaxies (UDGs) and compact dwarfs (CDs). Despite exhibiting stellar masses typical of dwarf galaxies, UDGs are characterized by incredibly large spatial extents that result in extremely low surface brightnesses.. They were recognized as a separate class of galaxies when a considerable number of such objects were discovered in the Coma cluster \citep{vanDokkum+2015ApJ...798L..45V}, although similar galaxies were studied previously \citep[e.g.,][]{Sandage+Binggeli1984AJ.....89..919S, Impey+1988ApJ...330..634I, Conselice+2003AJ....125...66C}. Subsequent observational efforts identified almost a thousand UDGs in the Coma cluster using Subaru deep imaging \citep{Koda+2015ApJ...807L...2K}. Since then, extensive samples in clusters and groups have established that UDGs are common across a wide range of halo masses and environments \citep{VanDerBurg+2017A&A...607A..79V}, with catalog-level work in the Coma cluster \citep{Yagi+2016ApJS..225...11Y, Bautista+2023ApJS..267...10B}, and more general catalogs \citep[e.g.,][]{Zaritsky+2022ApJS..261...11Z, Zaritsky+2023ApJS..267...27Z, Gannon+2024MNRAS.531.1856G} providing a widely used sample for comparison. Despite rapid observational progress, the physical nature of UDGs remains debated. Several formation mechanisms have been proposed, and many open questions remain regarding this galaxy population \citep[for a recent review, see][]{Gannon+2026}. 

From a theoretical perspective, cosmological and zoom-in simulations suggest that UDGs do not arise from a single formation channel but rather from a combination of internal processes and environmental effects. In isolation, diffuse stellar distributions can develop through strong stellar feedback that drives gas outflows and expands the stellar component \citep[e.g.,][]{DiCintio+2017MNRAS.466L...1D, Chan+2018MNRAS.478..906C}. In denser regions, such as galaxy clusters and groups, environmental mechanisms play a significant role. Numerical studies show that infalling dwarf galaxies can be transformed into extended systems through tidal heating and RPS during their interaction with the cluster potential \citep[e.g.,][]{Jiang+2019, Jackson+2021MNRAS.502.4262J}. Analyses of the IllustrisTNG simulations also support this picture. For example, \citet{Sales+2020MNRAS.494.1848S} demonstrated that the cluster UDG population in the TNG100 simulation is largely produced by tidal puffing up and structural expansion, while \citet{Benavides+2023MNRAS.522.1033B} showed that environmental processes play a significant role across a wide range of host halo masses. Similar conclusions emerged from the RomulusC zoom-in cluster simulation, where UDGs exhibit various formation histories but are ultimately shaped by the cluster environment and evolve into dispersion-dominated systems consistent with observations \citep{Tremmel+2019MNRAS.483.3336T, Tremmel+2020MNRAS.497.2786T}.

In stark contrast, CDs are some of the densest stellar systems known, bridging the gap between massive globular clusters and compact dwarf galaxies \citep{Norris+2014MNRAS.443.1151N, Wang+2023Natur.623..296W}. They were first identified in spectroscopic surveys of the Fornax cluster as a population of compact stellar systems with luminosities and sizes exceeding those of typical globular clusters, but significantly smaller than classical dwarf galaxies \citep{Hilker+1999A&AS..134...75H, Drinkwater+2000PASA...17..227D}. Since then, CDs have been detected in a wide variety of environments, including other galaxy clusters, groups, and even around isolated galaxies \citep[e.g.,][]{Mieske+2008A&A...487..921M, Brodie+2011AJ....142..199B, Norris+2011MNRAS.414..739N, Norris+2014MNRAS.443.1151N}. Nevertheless, they are most abundant in dense environments, particularly in the central regions of galaxy clusters, where they often appear to be strongly concentrated around massive galaxies.

The origin of CDs has also been debated, but some broad formation pathways are commonly discussed. In one scenario, CDs represent the high-mass extension of the globular cluster population, forming through the merging of massive star clusters during intense starburst events \citep{Fellhauer+Kroupa2002MNRAS.330..642F}. In the alternative scenario, CDs are the remnant nuclei of dwarf galaxies that have been tidally stripped during repeated interactions within the cluster or host galaxy potential, as proven by numerous studies \citep{bekki2001, bekki2003, pfeffer2013, pfeffer2014, martinovic2017, fm2018, kim2020, Deeley+2023MNRAS.525.1192D, Lohmann+2023MNRAS.524.5266L, Moura+2024MNRAS.528..353M}. Observational evidence supporting the stripping origin includes elevated dynamical mass-to-light ratios and the presence of supermassive black holes in some systems \citep[e.g.,][]{Seth+2014Natur.513..398S, Pfeffer+2016MNRAS.458.2492P}.

Although UDGs and CDs occupy opposite extremes of the structural parameter space, their cluster populations share several notable similarities. In particular, both types of galaxies are highly sensitive to environmental processes, and tidal interactions have been proposed as an important driver of their formation and evolution. Another intriguing clue linking UDGs and CDs arises from their spatial distributions within clusters. Observational studies have reported a pronounced anticorrelation between the two populations: CDs are predominantly found in dense cluster cores, whereas UDGs are largely absent from these regions and instead typically populate the cluster outskirts \citep{Janssens+2017ApJ...839L..17J, Janssens+2019ApJ...887...92J}. This pattern led to the hypothesis that at least a subset of CDs may originate from diffuse progenitors (UDGs) that were tidally transformed during their infall into clusters. Specifically, it was assumed that a nonnegligible fraction of UDGs host compact nuclei at the infall, which could plausibly survive severe subsequent tidal stripping, leaving behind compact remnants observable as CDs. However, the physical mechanism that would transform UDGs into CDs remains poorly understood from a theoretical point of view, particularly in a fully cosmological context.

In the first paper of the Puzzling Ultra-Diffuse Galaxy Evolution (PUDGE) series, we demonstrated that the IllustrisTNG\footnote{Publicly available at \url{https://www.tng-project.org/data/}.} (TNG100) simulation can host and reliably characterize even the most extreme UDGs, through a detailed case study of a simulated analog of the observed Nube galaxy \citep{Pavlov2025A&A...694A.312P}. In this paper, we test the evolutionary hypothesis (linking UDGs and present-day CDs in galaxy clusters) in a fully cosmological context using the IllustrisTNG simulation. To do so reliably, we first examine whether TNG100 reproduces the observed spatial anticorrelation between UDGs and CDs in galaxy clusters. We then utilize the time-domain information in the simulation to trace present-day CDs backward along their main progenitor branches, identifying systems that undergo at least a transient UDG phase during cluster infall. 

The paper is organized as follows. In Sect.~\ref{sec:methods}, we describe the simulation, sample selection, our definitions of the UDG and CD populations, and evolutionary analysis. In Sect.~\ref{sec:results}, we present the spatial distributions of UDGs and CDs in clusters at $z=0$ and the reconstructed evolutionary histories of the relevant progenitor systems. In Sect.~\ref{sec:discussion}, we discuss the physical interpretation of the results and place them in the context of the proposed formation scenario. Finally, we summarize our conclusions in Sect.~\ref{sec:summary}.

\section{Methods}\label{sec:methods}

We analyzed the formation and evolution of cluster galaxies using the TNG100 simulation, part of the IllustrisTNG magnetohydrodynamical suite \citep{Marinacci+2018, Naiman+2018, Nelson+2018, Pillepich+2018, Springel+2018}. The simulations were evolved using the moving-mesh code \texttt{Arepo} \citep{Springel2010AREPO}, incorporating a comprehensive galaxy formation model that includes radiative cooling, star formation, metal enrichment, and feedback from supernovae and supermassive black holes \citep{TNGmethods2017, TNGmethods2018}. The simulations adopt a Lambda cold dark matter cosmology consistent with \citet{PlanckColab+2016}, with matter density $\Omega_\mathrm{m} = 0.3089$, baryon density $\Omega_\mathrm{b} = 0.0486$, dark energy density $\Omega_\Lambda = 0.6911$, and Hubble constant $H_0 = 67.74\;\mathrm{km}\;\mathrm{s}^{-1}\;\mathrm{Mpc}^{-1}$. The evolution of the structure is traced from redshift $z = 127$ to the present day, with the outputs stored at 100 snapshots between $z = 20$ and $z = 0$.

Although the IllustrisTNG suite includes volumes of varying resolution and size \citep{Nelson+2019ComAC}, TNG100 was selected as the optimal laboratory for this study. The smaller TNG50 volume \citep{Nelson+2019MNRAS.490.3234N, Pillepich+2019MNRAS.490.3196P} offers a higher mass resolution; however, it contains only a single massive galaxy cluster, which is insufficient for a statistical analysis of environmental effects. In contrast, TNG100 offers a balance between volume and mass resolution, containing multiple galaxy clusters and, thus, enabling the study of cluster infall and tidal processing in dense environments. The mass resolution of TNG100 is $7.5 \times 10^{6}\;\mathrm{M}_\odot$ for dark matter particles and approximately $1.4 \times 10^{6}\;\mathrm{M}_\odot$ for baryonic particles. As demonstrated by \citet{Onions+2012MNRAS.423.1200O}, as well as in the first paper in this series \citep{Pavlov2025A&A...694A.312P}, this resolution is sufficient to reliably recover at least the global structural properties of dwarf galaxies, sufficient to identify the large-scale structural transformations investigated in this work. Moreover, \citet{Sales+2020MNRAS.494.1848S} explored the cluster population of UDGs in the TNG100 box, considering all galaxies with as few as 25 stellar particles. Similarly, several official TNG100 supplementary catalogs analyze galaxies in comparably low stellar mass regimes \citep[e.g.,][]{Genel+2015ApJ...804L..40G, Diemer+2019MNRAS.487.1529D}. We emphasize that the resolution limitations did not allow us to analyze the galaxies in detail (e.g., their internal or fine morphological structure), which is something we do not attempt to do and is out of the scope of the present study.

\subsection{Sample selection}\label{sec:sample}

Our initial analysis focuses on the final snapshot, which corresponds to the present-day redshift $z=0$. To begin, we construct a parent sample of all well-resolved genuine galaxies by selecting subhalos with \texttt{SubhaloFlag} $=1$\footnote{More information is available at \url{https://www.tng-project.org/data/docs/background} in the "Numerical Considerations and Issues" subsection; see also \citet{2025PASA...42...66M} for detailed discussion.}. From this parent sample, we restrict our analysis to galaxies within the stellar mass range $8 \leq \log (M_\star [\mathrm{M}_\odot]) \leq 10$. This range typically encompasses dwarf and intermediate-mass galaxies. Throughout this work, stellar, gas, and dark matter masses refer to the total subhalo quantities from the IllustrisTNG catalogs (\texttt{SubhaloMassType} parameter), as well as the total star formation rate (SFR) (\texttt{SubhaloSFR} parameter). Similarly, the stellar half-mass radius $R_{0.5,\star}$ corresponds to the catalog quantity \texttt{SubhaloHalfmassRadType} for the stellar component. Unless otherwise stated, all quantities are taken directly from the subhalo catalogs. We focus primarily on these global quantities because they are among the most robustly recovered galaxy properties in cosmological simulations for the galaxies in this mass range.

\begin{figure}[ht!]
\centering \includegraphics[width=\columnwidth, keepaspectratio]{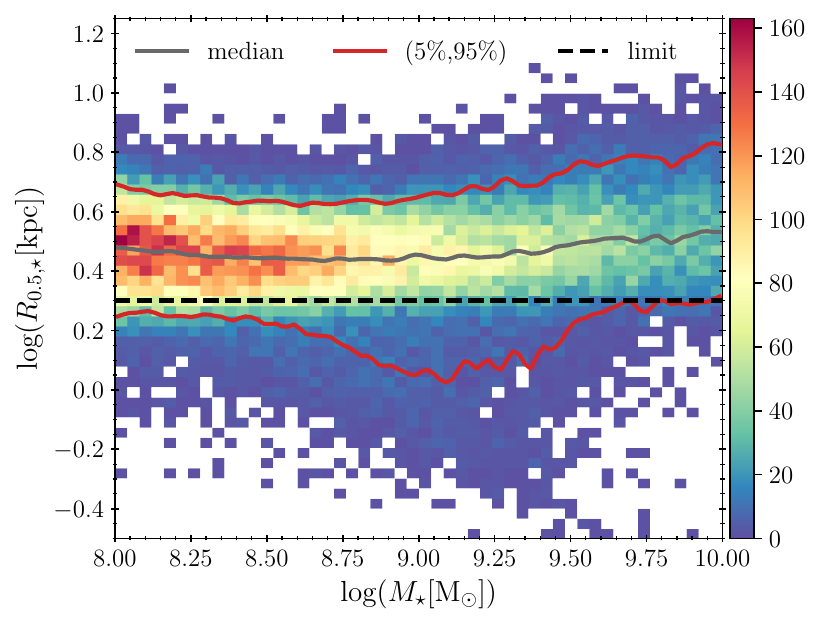}
\caption{Relevant part of the stellar mass-size relation for galaxies in the TNG100 simulation at $z=0$. The solid dark gray line shows the running median of the stellar half-mass radius $R_{0.5,\star}$, and the solid red lines represent the lower and upper fifth percentiles of the distribution. In contrast, the dashed black line represents our imposed limit for the CDs due to the excess of compact galaxies in this sample. The color bar represents the plain galaxy count.}
\label{fig:masssize_sample}
\end{figure}

To define our populations of interest, we first investigate the stellar mass–size relation for this sample, as presented in Fig.~\ref{fig:masssize_sample}. We calculate the running median of the size distribution as a function of mass, along with the fifth and 95th percentiles, to delineate the loci of typical and outlier galaxy populations. For selecting UDGs, we adopt a purely structural definition based on their extreme size for a given stellar mass. Specifically, we classify all galaxies lying above the 95th percentile of the mass-size relation as UDGs. This approach, which selects for the most spatially extended galaxies, is consistent with methodologies used in previous studies of UDGs in cosmological simulations, particularly that of \citet{Benavides+2023MNRAS.522.1033B}. This enables a direct theory-to-theory comparison by avoiding the complexities of mocking observational surface-brightness criteria.

For selecting CDs, we originally intended to use a similar statistical criterion. However, we can see a distinct asymmetry in the mass-size relation shown in Fig.~\ref{fig:masssize_sample}, that is, there is a significant excess of compact galaxies which affects the fifth percentile line, particularly at stellar masses around or greater than $\log (M_\star [\mathrm{M}_\odot]) = 9$. Since we did not separate central and satellite galaxies in our sample, this feature is a physically motivated outcome of galaxy evolution in the simulation, reflecting the large population of satellite galaxies whose sizes have been dramatically reduced by tidal effects. Given this physical pile-up of compact systems, a simple percentile cut does not yield a clean sample. Instead, we adopt a more robust, physically motivated criterion, defining CDs as all galaxies in our mass range with a stellar half-mass radius $R_{0.5,\star}< 2\; \mathrm{kpc}$. This cutoff roughly corresponds to the fifth percentile line in both the lower- and higher-mass regimes, while providing a clear definition that selects the most compact systems across the full mass range. Additionally, the adopted threshold ensures that the systems, analyzed here in detail, remain comfortably above the nominal TNG100 softening scale (see Table~\ref{tab:ucd_properties}). The softening length in the TNG100 box for the collisionless components (dark matter and stellar particles) is approximately $\epsilon \approx 0.74\;\mathrm{kpc}$, corresponding to the limit of roughly $\log(R_{0.5,\star}[\mathrm{kpc}]) \simeq -0.13$, meaning that a small portion of the excess population of CDs visible in Fig.~\ref{fig:masssize_sample} is not resolved above this scale. We excluded this subpopulation from further analysis. The adaptive softening for gas cells can be as low as $\epsilon_\mathrm{gas} \simeq 0.125\;\mathrm{ckpc}/h$, though this is not the relevant scale for the stellar structural quantities analyzed here. We also exclude spurious subhalos that form late in the simulation (i.e., in a few last snapshots).

After splitting the sample into UDGs and CDs, we had to determine their environment and, more specifically, associate individual galaxies with their host halos that satisfy the present-day criterion for galaxy clusters \citep[given by][]{Paul+2017}. A host halo is classified as a cluster if its total mass exceeds $8 \times 10^{13}\;\mathrm{M_{\odot}}$. Additionally, to ensure the robustness of our results and avoid numerical artifacts associated with periodic boundary conditions, we restrict our analysis to galaxy clusters located within the central region of the simulation volume. As demonstrated by \citet{Mitrasinovic2025A&A...703L..16M}, the IllustrisTNG simulations contain a population of spurious objects near the edges of the simulation box that mimic the properties of tidally stripped galaxies. It was specifically highlighted that TNG100 features massive galaxy clusters bisected by simulation boundaries, where numerical artifacts are conflated with genuine galaxies affected by environmental processing. Following the recommendation for statistical studies, we exclude all clusters situated within the buffer zone near the box boundaries to ensure that the evolutionary tracks and spatial distributions derived in this work are driven solely by physical processes. Applying these spatial constraints yields a final, uncontaminated sample of seven galaxy clusters (\texttt{GrNr}\footnote{This parameter refers to the parent Friends-of-Friends (FoF) halo unique identifier in the FoF Halos catalog.} = 1, 2, 8, 9, 11, 14, and 18 at $z=0$). All subsequent analysis is restricted to the UDG and CD populations hosted within these systems. We also emphasize that, although we consider UDG and CD populations within clusters, not all of those galaxies are classified as satellites, despite their cluster association.

\subsection{Historical analysis of the CD population and identification of UDG progenitors}\label{sec:8candidates}

To test the hypothesis that CDs are the tidally stripped remnants of UDGs, we trace the evolutionary histories of all $112$ CDs identified at redshift $z=0$ within the seven selected clusters. For each CD in our sample, we follow its main progenitor across snapshots using the \texttt{SubLink} merger trees \citep{Rodriguez-Gomez+2015MNRAS.449...49R} and examine the temporal evolution of its key structural properties (i.e., the stellar, gas, and dark matter masses, the stellar half-mass radius, the SFR, and environment (i.e., its host halo properties and the 3D distance relative to the center of its host, cluster or proto-cluster, halo).

A progenitor is classified as a viable UDG precursor only if it satisfies three criteria. First, the galaxy must remain in the UDG regime for at least multiple consecutive snapshots, ensuring that the extended phase reflects a genuine structural state rather than a short-lived fluctuation or artificial size enhancement. Second, this diffuse phase must occur immediately prior to, or at the time of, infall into a galaxy cluster or proto-cluster at earlier epochs, defined as crossing the host halo virial radius. This makes it more likely that the enhanced size is physically associated with environmental processing, and not artificially induced. Third, the system must subsequently exhibit clear signatures of tidal stripping and mass loss, consistent with environmental processing within the cluster potential. These criteria isolate systems in which the diffuse phase is dynamically linked to environmental transformation, rather than internal secular evolution or early-time structural fluctuations. Applying these constraints, we identify a subset of eight CDs whose evolutionary histories are consistent with a transient UDG phase followed by intense tidal processing within the cluster environment. Therefore, the selected systems do not correspond to isolated single-snapshot excursions in stellar size.

Because generating and statistically evaluating the complete stellar mass–size relation at every individual snapshot is computationally prohibitive, a progenitor galaxy is classified as UDG if it satisfies the same structural criterion adopted at redshift $z=0$, that is, lying above the 95th percentile of the present-day relation defined in Sect.~\ref{sec:sample}. Although the mass-size relation evolves significantly with redshift, where both observed and simulated galaxies were systematically more compact at earlier epochs at fixed stellar mass \citep[e.g.,][]{Trujillo+2006ApJ...650...18T, VanDerWel+2014ApJ...788...28V, Shibuya+2015ApJS..219...15S, Genel+2018MNRAS.474.3976G}, we apply the present-day relation uniformly across all snapshots. This choice ensures a conservative selection. Galaxies identified as UDGs at earlier times represent the most extreme outliers relative to the evolving population, providing a strict lower limit on the fraction of CDs with diffuse progenitors.

\section{Results}\label{sec:results}

\begin{figure}[ht!]
\centering \includegraphics[width=\columnwidth, keepaspectratio]{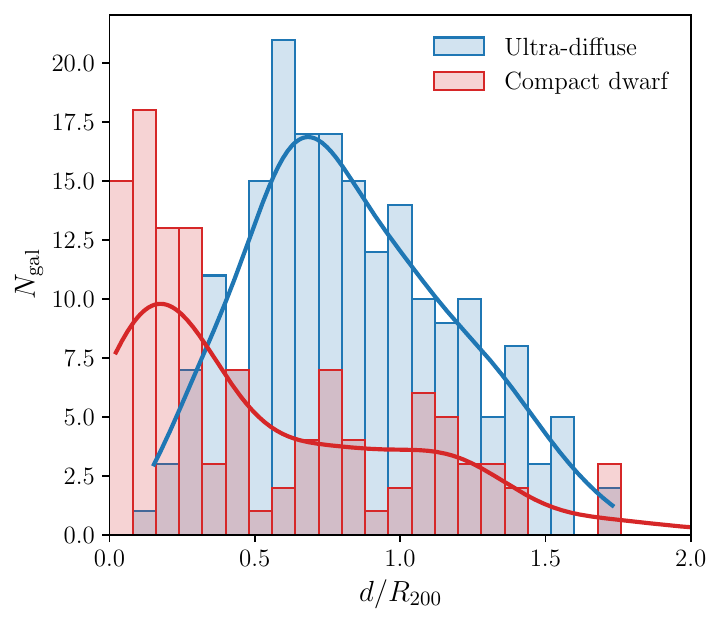}
\caption{Spatial distributions of the UDG (blue histogram and line) and CD (red histogram and line) populations within their host galaxy clusters at $z=0.$ The $x$-axis shows the 3D cluster-centric distance normalized by the virial radius of the host halo ($d/R_{200}$). The $y$-axis represents the simple number count (histograms) with overplotted probability density function (lines) for each sample.}
\label{fig:selectedclusters}
\end{figure}

As mentioned previously, recent observational studies have revealed a striking spatial anticorrelation between UDGs and CDs in galaxy clusters \citep{Janssens+2017ApJ...839L..17J, Janssens+2019ApJ...887...92J}. Our first objective is to determine whether this observed phenomenon is reproduced in the TNG100 simulation, thereby validating the simulation as a reliable tool for studying its potential evolutionary connection. In Fig.~\ref{fig:selectedclusters}, we present the 3D spatial distribution of our selected UDG and CD samples within their host clusters. The cluster-centric distance is normalized by the virial radius of each host halo to allow for a robust statistical comparison between clusters of varying sizes. There is a clear and strong spatial anticorrelation between the two populations, in excellent agreement with the observational findings. The CDs are highly centrally concentrated, with their probability density peaking in the innermost cluster core, well within $d/R_{200} \simeq 0.2$. In contrast, the UDGs preferentially inhabit the cluster outskirts. Their distribution is broadly peaked around $d/R_{200} \simeq 0.7$, extending well beyond the virial radii of the clusters, and exhibits a sharp decline in the cluster centers, where the CD population is dominant.

\begin{figure}[ht!]
\centering \includegraphics[width=\columnwidth, keepaspectratio]{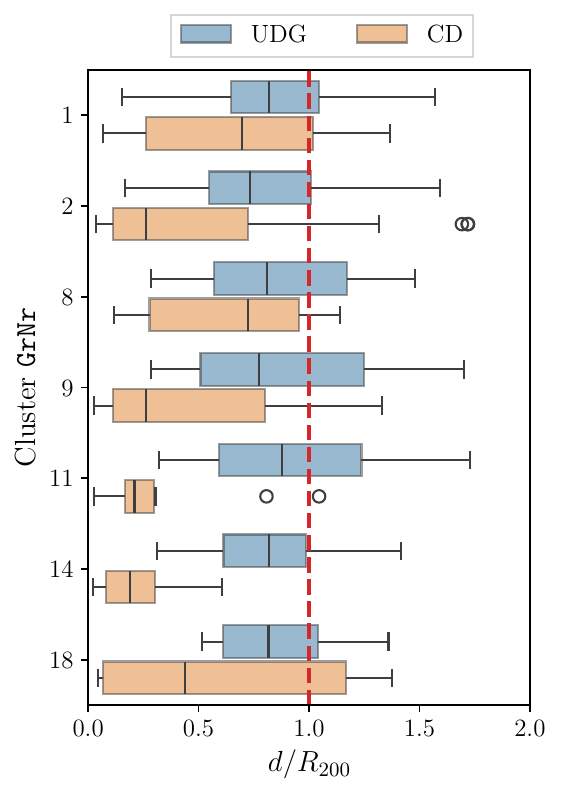}
\caption{Similar to Fig.~\ref{fig:selectedclusters}, the spatial distribution of the UDG (blue) and CD (orange) populations is shown but separately within each individual cluster. The vertical red dashed line represents the virial radius of each cluster. The $y$-axis is labeled with unique cluster IDs, called \texttt{GrNr}, at redshift $z=0$, i.e., snapshot $99$ in the simulation.}
\label{fig:percluster}
\end{figure}

To confirm that this anticorrelation is not merely an artifact of stacking multiple clusters, but rather a pervasive feature of individual cluster environments, we also examine the spatial distributions in each cluster. Fig.~\ref{fig:percluster} presents box-and-whisker plots of the UDG and CD cluster-centric distances for every individual galaxy cluster. Each box shows the interquartile range of distances, with the median indicated by the central line. Consistently in all displayed clusters, to varying degrees, the CDs (orange boxes) are found to be significantly closer to the cluster center, with their median distances typically well below $d/R_{200}=0.5$. In contrast, UDGs (blue boxes) remain consistently farther out, with their median distances often falling between $d/R_{200}=0.5$ and $d/R_{200}=1$, and extending beyond the virial radius of a cluster. This confirms that the observed spatial anticorrelation is a robust property present within individual clusters in the TNG100 simulation.

Having established this consistency with observational studies, we can now leverage the unique power of the simulation to move beyond this static snapshot at redshift $z=0$. We traced the CD population back in time to directly test the hypothesis that they are tidally stripped remnants of infalling UDGs with prominent cores.

\subsection{UDG progenitors of present-day CDs in galaxy clusters}

To reconstruct the evolutionary pathways of the eight selected CD candidates with UDG progenitors (identified as described in Sect.~\ref{sec:8candidates}), we tracked their defining parameters as a function of the lookback time. Specifically, we monitored their mass components (dark matter, stars, and gas), SFR, and stellar size, alongside the host halo mass and halo-centric (or cluster-centric) distances for each candidate (two illustrative examples on Fig.~\ref{fig:fourplots} (see Appendix~\ref{sec:candidates} for the rest of the candidates)). Here, host halo mass $M_\mathrm{host}$ denotes the total mass of the parent FoF halo associated with the galaxy at a given snapshot, while $R_{200}$ represents the corresponding virial radius enclosing a mean density 200 times the critical density of the Universe\footnote{Corresponding to the IllustrisTNG FoF Halos catalog quantity \texttt{Group\_R\_Crit200}.}. Tracking the infall and repeated spiraling of the UDG progenitors around the host halo reveals a remarkably consistent evolutionary sequence for most galaxies after they enter the host environment. In some cases, apparent jumps in the relative cluster-centric radius $d/R_{200}$ reflect changes in the host halo center and virial radius during host halo mergers, which are simultaneously reflected by instantaneous increases in the host halo mass, rather than from abrupt physical displacement of the galaxy itself.

\begin{figure*}[ht!]
    \centering
    \newcommand{\subfigwidth}{0.48\textwidth}

    \begin{subfigure}{\subfigwidth}
        \includegraphics[width=0.97\columnwidth]{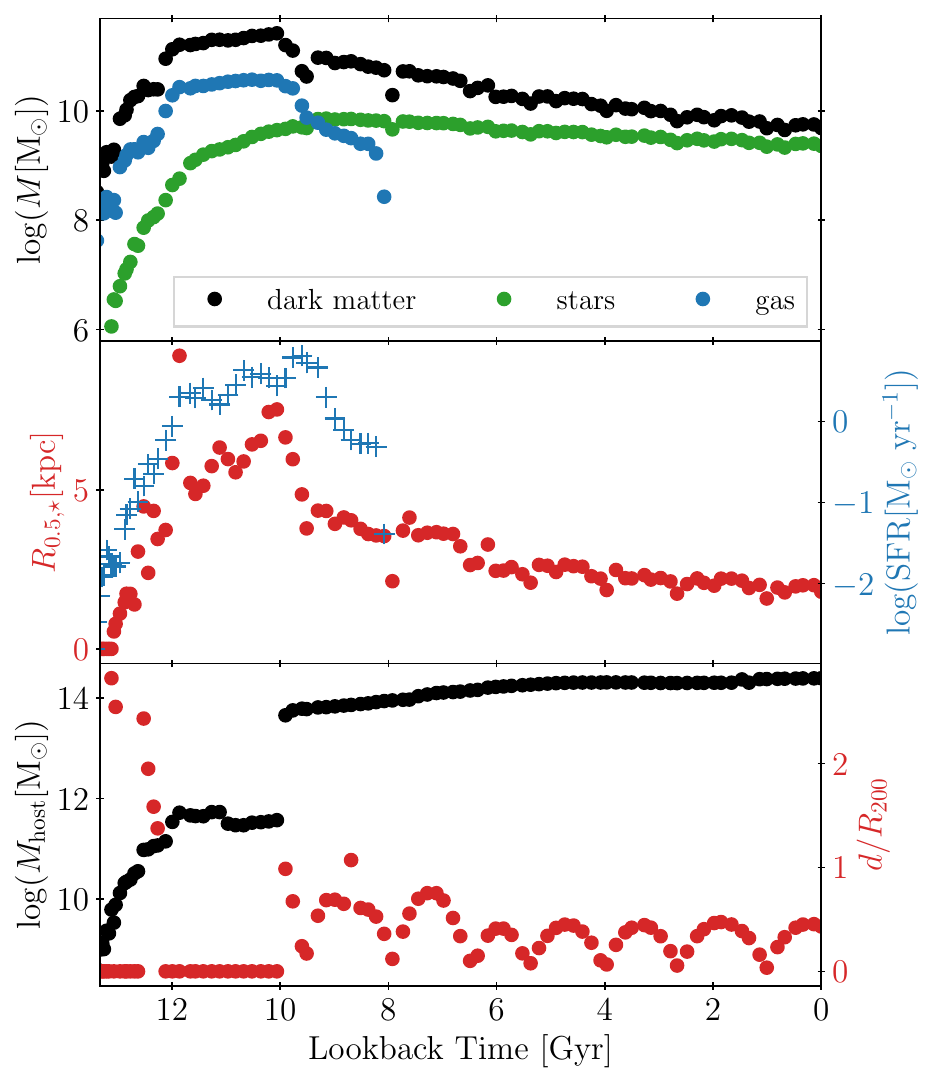}
        \caption{ID83402}
        \label{fig:83402}
    \end{subfigure}
    \hfill 
    \begin{subfigure}{\subfigwidth}
        \includegraphics[width=0.97\columnwidth]{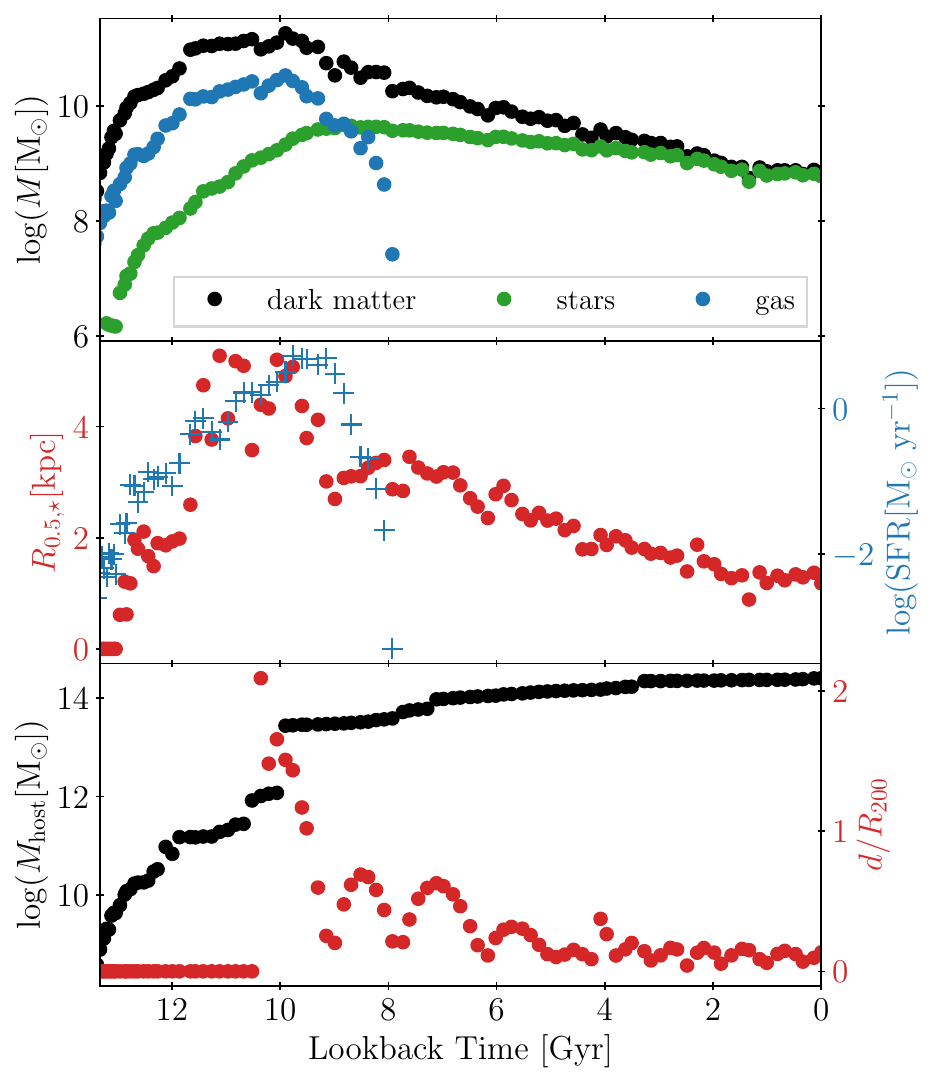}
        \caption{ID89140}
        \label{fig:89140}
    \end{subfigure}

    \caption{Evolutionary histories of two illustrative examples of CD progenitors that underwent a transient UDG phase (see Appendix~\ref{sec:candidates} for the rest of the candidates). Each subfigure represents a different galaxy, indicated by its present-day \texttt{SubfindID}. Top panels: Mass evolution of the individual components of the galaxy (dark matter, stars, and gas). Middle panels: Evolution of the stellar half-mass radius (red, left $y$-axis) and SFR (blue, right $y$-axis). Bottom panels: Evolution of the host halo mass that represents the mass of the cluster at the present day (black, left $y$-axis) and relative cluster-centric radius (red, right $y$-axis).}
    \label{fig:fourplots}
\end{figure*}

\begin{figure}[ht!]
\centering \includegraphics[width=\columnwidth, keepaspectratio]{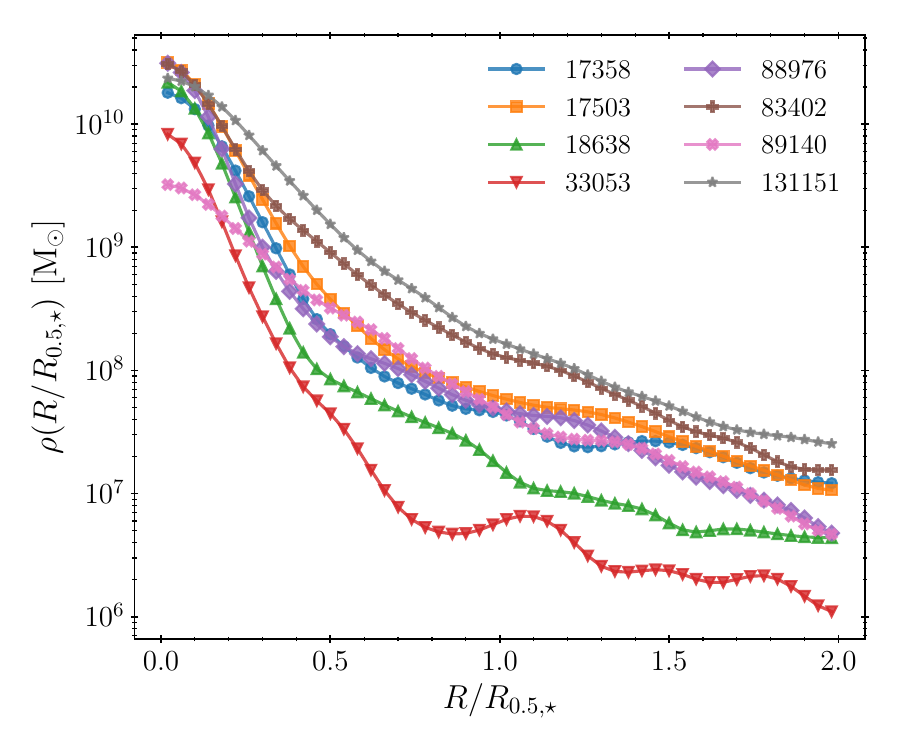}
\caption{Stellar mass density profiles as a function of relative radii (scaled by the stellar half-mass radius $R_{0.5, \star}$) for each UDG candidate at maximum radii redshift, labeled by present-day \texttt{SubfindID}.}
\label{fig:udgdensities}
\end{figure}

During this orbital phase, the masses of dark matter and stars (to some extent) present during the most extended UDG phase (at a lookback time of $\sim 10$ Gyr for most candidates) exhibit a steady, continuous decline. This gradual depletion of material from the outskirts points directly to relentless tidal stripping driven by the host halo. However, an examination of the stellar mass density profiles of these corresponding UDG phases (Fig.~\ref{fig:udgdensities}) reveals an apparent inconsistency: none of the candidates exhibit a distinct, preexisting central core. While the resolution did not allow us to fully resolve nuclear-scale structures, the absence of any significant central excess in the stellar density profiles suggests that no dominant preexisting core is present at the resolved scales. Given this absence of an early core, combined with the continuous tidal stripping of their outer layers, the resulting CDs could not have formed early on and simply been left behind as exposed remnants.

The mechanism resolving this apparent contradiction is revealed by the gas reservoirs of galaxies. Concurrently with the stripping of the outskirts, the middle panels show a rapid spike in the SFR, indicating that the bulk of the available gas was either violently consumed by star formation or stripped away. Simultaneously, the stellar half-mass radius is seen transitioning from typical UDG values, ultimately settling into compact dimensions characteristic of CDs. This sequence suggests an unexpected formation pathway: the central cores that eventually become CDs were forged entirely later in the timeline. Environmental disturbances essentially squeezed the galaxy, concentrating at least a part of the existing gas reservoir toward the center and fueling a subsequent burst of star formation just as the outer material was being torn away.

Crucially, inspection of Fig.~\ref{fig:fourplots}, Fig.~\ref{fig:fourplots+}, and Fig.~\ref{fig:fourplots++} reveals that the peak SFR during this transformation window is the highest each of these candidate galaxies achieves across their entire lifetimes. The cluster environment therefore does not merely reactivate modest star formation in a system undergoing disruption and transformation, but instead drives the single most intense starburst event in the history of the galaxy. It is particularly interesting to point out that despite the presence of such intense star formation episode, the stellar mass evolution does not exhibit a corresponding increase. Instead, the stellar mass either declines or remains approximately constant relative to its pre-infall value for all of the candidates (see top panels in Fig.~\ref{fig:fourplots}, Fig.~\ref{fig:fourplots+}, and Fig.~\ref{fig:fourplots++}). This implies that tidal stripping of the stellar component is highly efficient and, in most cases, outweighs the stellar mass formed during the burst. As a result, the net stellar mass evolution alone underestimates the degree to which the stellar component is being reshaped during this transformation.

To quantitatively confirm whether these present-day galaxies are truly a synthesized product of stripped outskirts and newly formed stars, we analyzed the mass fraction of stars formed after the UDG phase ($f_{\rm new}$), detailed in Table~\ref{tab:ucd_properties}. This quantity is computed using particle-level information extracted from raw simulation cutouts for each galaxy. Specifically, each stellar particle in the simulation carries a formation scale factor, which records the cosmic time at which it was created\footnote{This parameter is called \texttt{GFM\_StellarFormationTime} in the raw cutout, or \texttt{aform} if the cutout is loaded using \texttt{pynbody} python package.}. For each galaxy, we identified the snapshot corresponding to its maximum stellar half-mass radius and used it as a reference epoch. We then calculated the mass-weighted fraction of present-day stellar particles whose formation times correspond to later epochs. The resulting values provide a clear and robust measure of stellar mass assembly. In all cases, the majority of the present-day stellar mass is formed after the galaxy reaches its most extended (i.e., UDG) phase, demonstrating that the compact stellar systems observed at $z=0$ are largely assembled during the transformation rather than inherited from the UDG phase. The epoch of maximum stellar half-mass radius is used as a representative reference point within the sustained diffuse evolutionary phase, rather than as the sole basis for UDG classification.

The limits of this mechanism are evident in the single exception in our sample, the candidate ID88976 (see Fig.~\ref{fig:88976}). This galaxy reaches its most extended UDG phase somewhat later - upon entering the host halo at a lookback time of around 8 Gyr. Because of this delayed evolution, it exhibits a more gradual SFR with the lowest SFR peak and a longer-lasting gas reservoir. Although earlier star formation had already depleted some of its total gas content (resulting in a lower $f_{\rm new}$ compared to other candidates), enough gas remained to trigger late-stage core formation, leaving the final galaxy still largely composed of newly formed stars. Ultimately, this outlier confirms the overarching physical reality of this UDG-to-CD pathway: it is a highly efficient transformative process, provided there is sufficient gas present to fuel the creation of the central core. Consequently, this mechanism must have been vastly more probable in the gas-rich environments of the early universe than at later epochs.

\begin{table*}
\caption{Properties of the eight CD candidates whose progenitors underwent a transient UDG phase prior to cluster infall.}
\label{tab:ucd_properties}
\centering
\begin{tabular}{c c c c c c c c c c c}

SubfindID &
$t_{\rm UDG}$ &
$\log(M_{\star,\rm UDG})$ &
$R_{0.5,\star}^{\rm UDG}$ &
$f_{\rm gas,UDG}$ &
$\log(M_{\star,\rm CD})$ &
$R_{0.5,\star}^{\rm CD}$ &
${\rm SFR_{peak}}$ &
$f_{\rm new}$ &
$N^\mathrm{CD}_\mathrm{part,\star}$ &
\texttt{GrNr} ($z$=0)\\
~ &
[Gyr] &
[$\rm M_{\odot}$] &
[kpc] &
~ &
[$\rm M_{\odot}$] &
[kpc] &
[$\rm M_{\odot}\,yr^{-1}$] &
~ &
~ & 
  \\
\hline
17358  & 11.26 & 9.17 & 10.58 & 0.966 & 9.73 & 1.68 & 22.42 & 0.919 & 5930 & 1 \\
17503  & 10.82 & 9.35 &  7.42 & 0.941 & 9.36 & 1.39 & 12.06 & 0.863 & 2514 & 1 \\
18638  & 11.26 & 8.88 &  9.89 & 0.965 & 8.48 & 1.36 &  4.83 & 0.925 & 334  & 1 \\
33053  & 11.42 & 8.39 & 10.79 & 0.985 & 8.14 & 1.96 &  3.13 & 0.964 & 155  & 2 \\
83402  & 10.21 & 9.62 &  7.47 & 0.898 & 9.36 & 1.80 &  6.45 & 0.763 & 2504 & 8 \\
88976  &  8.08 & 9.12 &  6.67 & 0.913 & 8.91 & 1.85 &  1.48 & 0.653 & 899  & 9 \\
89140  & 10.06 & 9.23 &  5.20 & 0.943 & 8.77 & 1.18 &  5.31 & 0.915 & 683  & 9 \\
131151 & 10.06 & 9.89 &  5.00 & 0.794 & 9.33 & 0.87 & 17.70 & 0.839 & 2344 & 18 \\
\hline
\end{tabular}
\tablefoot{
SubfindID refers to the present-day ($z = 0$) identifier in the TNG100 simulation.
$t_{\rm UDG}$ is the lookback time at the epoch of maximum stellar half-mass radius,
corresponding to redshifts in the range $z \approx 1.0$--$2.7$.
$\log(M_{\star,\rm UDG})$ and $R_{0.5,\star}^{\rm UDG}$ are the stellar mass and
stellar half-mass radius at this epoch.
$f_{\rm gas,UDG}$ is the gas mass fraction at the same epoch, defined as
$M_{\rm gas} / (M_{\rm gas} + M_{\star})$.
$\log(M_{\star,\rm CD})$ and $R_{0.5,\star}^{\rm CD}$ are the present-day stellar
mass and half-mass radius.
${\rm SFR_{peak}}$ is the peak SFR across the full lifetime of each system,
which in all cases occurs during the environmental transformation window.
$f_{\rm new}$ is the mass-weighted fraction of present-day stellar mass formed after
the epoch of maximum extent. $N^\mathrm{CD}_\mathrm{part,\star}$ represents the total number of stellar particles in the present-day CD galaxy, where all sit comfortably above the limit of $25$ stellar particles adopted by \citet{Sales+2020MNRAS.494.1848S}. \texttt{GrNr} indicates the present--day host cluster ID of the CD candidates.} 

\end{table*}

Notably, two of the eight candidates, IDs 18638 and 33053 (with 334 and 155 stellar particles), respectively, are the least resolved systems in the sample (see Table~\ref{tab:ucd_properties}). Although both surpass the threshold of $25$ stellar particles adopted by \citet{Sales+2020MNRAS.494.1848S} for the study of UDG populations in the same TNG100 simulation, a larger number of particles is generally desirable for robust recovery of structural properties such as the stellar half-mass radius and the stellar mass density profile. We therefore caution that the quantitative values reported for these two systems in Table~\ref{tab:ucd_properties} should be interpreted with greater care than those of the remaining six candidates. Nevertheless, their qualitative evolutionary behavior is fully consistent with the rest of the sample: both exhibit extremely high gas fractions during their UDG phase, a pronounced SFR spike coinciding with environmental infall, a clear transition from extended to compact dimensions, and high fractions of newly formed stellar mass. Restricting the analysis to the six better-resolved candidates does not alter the proposed formation pathway or any of the main conclusions of this work.

The host cluster ID (\texttt{GrNr} at z = 0) of each candidate is listed in Table~\ref{tab:ucd_properties}. The candidates are distributed across five of the seven clusters; notably, none reside in the two clusters (\texttt{GrNr} 11 and 14) exhibiting the minimal overlap of UDG and CD spatial distributions. We note, however, that the present--day host cluster does not directly trace the site of transformation, since the progenitor UDG phase and infall occurred at earlier epochs.

Taken together, the absence of a preexisting stellar core in the UDG progenitors (Fig.~\ref{fig:udgdensities}), the consistent pattern of tidal mass loss accompanied by an intense centrally concentrated starburst (Fig.~\ref{fig:fourplots},  Fig.~\ref{fig:fourplots+}, and Fig.~\ref{fig:fourplots++}), and the high fractions of newly formed stellar mass (Table~\ref{tab:ucd_properties}) paint a coherent and physically motivated picture. The compact stellar component of the present-day CD in each of these eight systems is not the remnant nucleus that survived tidal stripping, but a newly assembled object built from the compressed gas reservoir of an extremely diffuse, gas-rich progenitor during its transformation within the cluster environment. The physical interpretation of this result and its implications for the broader UDG and CD populations are discussed in Sect.~\ref{sec:discussion}.

\section{Discussion}\label{sec:discussion}

The results presented in Sect.~\ref{sec:results} establish two complementary findings. First, the TNG100 simulation reproduces the observed spatial anticorrelation between UDGs and CDs in galaxy clusters \citep{Janssens+2017ApJ...839L..17J, Janssens+2019ApJ...887...92J}, validating its use as a laboratory for studying the evolutionary connection between these populations. Second, tracing the complete histories of 112 cluster CDs reveals a subset of eight systems whose progenitors pass through a genuine UDG phase immediately prior to cluster infall, undergo a compression-driven starburst that constitutes the most intense star-forming episode in their lifetimes, and assemble the overwhelming majority of their present-day stellar mass after the epoch of maximum spatial extent.

The classical scenario for CD formation via tidal stripping posits that a compact stellar nucleus must be present in the infalling progenitor prior to environmental processing, with the nucleus surviving as the stripped remnant once the surrounding stellar envelope has been removed \citep[e.g.,][]{bekki2001, bekki2003, pfeffer2013, pfeffer2014}. The results presented here are fundamentally inconsistent with this picture, as none of the progenitors of the eight identified candidates exhibit a distinct central concentration at the UDG epoch (which roughly corresponds to the time of the infall and subsequent environmental processing). Instead, a majority of the present-day stellar mass in each system was assembled after the epoch of maximum spatial extent, and therefore each resulting CD is not an exposed preexisting nucleus but a newly assembled object, built from the compressed gas reservoir of the infalling diffuse progenitor during the transformation process itself.

The physical sequence underlying this process is consistent across all candidates, as described in detail in Sect.~\ref{sec:results}. The transformation therefore proceeds through a coupled process (a combination of various effects and mechanisms and an intricate chain of events) in which the diffuse galaxy is simultaneously destroyed at large radii and rebuilt at its center, producing a compact remnant whose stellar mass is dominated by stars formed during this phase. This mechanism shares similarities with the burst-driven formation channel for compact stellar systems identified by \citet{Bian+2025ApJ...979L..33B}, in which ram-pressure-induced gas compression triggers rapid star formation during pericentric passages in TNG50. However, our results place this process in a distinct evolutionary context, and the two scenarios differ in important respects. In \citet{Bian+2025ApJ...979L..33B}, the starburst is driven primarily by short, high-velocity pericentric passages at small galactocentric distances, with ram pressure playing the dominant compressive role and metal enrichment a central focus. Moreover, their initial sample selects compact stellar systems mainly on the basis of the fact that they are metal-rich outliers of the mass-metallicity relation \citep{Gallazzi+2005MNRAS.362...41G, Kirby+2013ApJ...779..102K}. In contrast to \citep{Bian+2025ApJ...979L..33B}, our sample selection does not employ metallicity as a filtering criterion, but is instead based on structural properties and cluster affiliation. Also, the transformation here unfolds over a longer timescale through repeated spiraling within the cluster potential, and the progenitors are among the most extreme galaxies in the simulation: not just structurally most extended, but also extremely gas-rich. The two channels may therefore represent distinct physical realizations of a more general class of environmentally driven compact system formation, operating under different conditions of orbital geometry, environmental density, and progenitor gas content. 

One of the key implications of our results is that UDGs do not necessarily represent a stable or long-lived class of galaxies. In fact, this structural state in the systems examined here is a transient phase reached shortly before or during cluster infall, after which environmental processing drives rapid structural transformation into a remnant CD. This is consistent with the broader picture in which UDGs constitute a heterogeneous population shaped by both internal processes and environmental effects \citep[e.g.,][]{Sales+2020MNRAS.494.1848S, Tremmel+2020MNRAS.497.2786T, Benavides+2023MNRAS.522.1033B, Buzzo+2025MNRAS.536.2536B, Gannon+2026}, and adds a specific and physically motivated exit channel to this picture. Thus, the pathway identified here provides a physically motivated evolutionary channel through which diffuse galaxies can evolve into compact stellar systems.

The efficiency of this UDG-to-CD transformation is critically dependent on the gas content of the progenitor at the infall. The systems identified in this work are extremely gas-rich and are accreted at relatively early cosmic times, when such conditions were common among infalling field galaxies \citep{Tacconi+2020ARA&A..58..157T, Walter+2020ApJ...902..111W}. The boundary conditions of the transformation channel are illustrated by the candidate ID88976 (Fig.~\ref{fig:88976}), which infalls the latest and has the lowest gas fraction and the peak SFR in the sample, resulting in the lowest fraction of the newly formed stellar mass. Therefore, it is unlikely that this channel will operate for the majority current UDGs in clusters, which are typically gas-poor as a consequence of pre-processing in group environments \citep{Reynolds2022, Holwerda2025}, or RPS at the cluster periphery prior to virial radius crossing \citep{Lopes+2024MNRAS.527L..19L, Piraino-Cerda+2024MNRAS.528..919P, Xie+2025A&A...698A..73X}. For example, the UDG candidate VCC 1964 is currently undergoing RPS disruption as it enters the cluster \citep{Taylor+2026A&A...707A.107T}. Some of those cluster UDGs, as they spiral deeper into the cluster potential, are expected to experience even more extreme environmental processing that would eventually cause significant mass loss without a compensating starburst \citep[e.g.,][]{Junais+2021A&A...650A..99J, Junais+2022A&A...667A..76J}, ultimately dissolving them into the intracluster stellar field rather than producing a compact remnant \citep[e.g.,][]{Roman+2021A&A...649L..14R}. An exception would be those rare present-day infalling UDGs that already host a compact preexisting stellar nucleus, which could plausibly survive stripping through the classical channel. Another possibility is the existence of a distinct subset of UDGs at large cluster-centric radii that may avoid transformation altogether. This can happen if they remain on high-energy orbits that only graze the cluster periphery, experiencing sufficiently weak environmental effects to survive as extended, low-surface brightness systems for extended periods \citep[so-called backsplash galaxies or the ones escaping cluster environment; see, e.g.,][]{Gill+2005MNRAS.356.1327G, Ludlow+2009ApJ...692..931L, Benavides+2021NatAs...5.1255B, Borrow+2023MNRAS.520..649B, Mitrasinovic+2023A&A...680L...1M, Paudel+2025A&A...701L...9P, Smole+2025PASA...42...86S}. 

Of the 112 CDs examined here, only eight meet the criteria for the UDG-to-CD transformation, corresponding to approximately 7 per cent of the sample. This should be regarded as a conservative lower limit, given the strict classification criteria and the use of a fixed present-day structural definition at all redshifts, which identifies only the most extreme structural outliers at earlier epochs. The remaining systems do not represent a uniform population: some may follow the same starburst-driven channel identified here but failed to satisfy the strict UDG phase criteria, while the broader sample encompasses galaxies formed through the classical tidal stripping of nucleated progenitors after cluster infall, as well as systems that formed entirely in situ \citep[e.g.,][]{martinovic2017, Deeley+2023MNRAS.525.1192D}. The relative contribution of each of these formation channels to the overall cluster CD population, and whether the starburst-driven pathway extends more broadly to gas-rich diffuse progenitors that do not formally qualify as UDGs, remain open questions that will be addressed in future work.

Finally, we note that the quantitative details of the transformation are subject to the limitations of TNG100, including finite mass resolution and subgrid modeling of star formation and feedback \citep{TNGmethods2018}. Although these factors may affect the precise values of stellar mass and SFR, for example, they are unlikely to alter the qualitative picture, which is governed by large-scale environmental processes that are well resolved. The UDG-to-CD transformation channel identified here should therefore be regarded as physically plausible and supported by a consistent set of independent diagnostics across all eight systems, while its overall quantitative contribution to the cluster CD population remains to be established.

\section{Conclusions}\label{sec:summary}

In this work, we have utilized the TNG100 cosmological simulation to investigate the proposed evolutionary link between UDGs and CDs in the extreme environments of galaxy clusters. Our primary findings are summarized as follows:

\begin{itemize}
    \item We demonstrate that TNG100 successfully reproduces the observed spatial anticorrelation between the two populations. Present-day CDs are highly centrally concentrated within the cluster cores ($d/R_{200} \lesssim 0.2$), while UDGs are preferentially found in the cluster outskirts, validating the simulation as a robust tool for studying this environmental transformation.
    
    \item By tracing the main progenitor branches of 112 cluster CDs back to $z \approx 3$, we identified a distinct and substantial subset of about 7\% of systems that underwent a transient UDG phase immediately prior to cluster infall indicating a viable route of their formation. 
    
    \item We find that the progenitors of these CDs lacked prominent stellar nuclei during their diffuse phase. Instead, the compact stellar component of the resulting CD is freshly assembled during the infall process. Environmental interactions drive a compression-driven starburst, the most intense in the history of each galaxy, concentrating the gas reservoir into a new, dense stellar core.
    
    \item This transformation pathway is characterized by high mass-weighted fractions of newly formed stars ($f_{\rm new} \gtrsim 0.65$--$0.96$). The resulting CD is thus a synthesized product of tidal stripping and active star formation, rather than an exposed, preexisting remnant.
    
    \item The high gas richness ($f_{\rm gas} \gtrsim 0.8$) at the time of infall is a critical prerequisite for this channel. This suggests that while this pathway may be less common in the quenched populations of the local Universe, it was likely a dominant mechanism for CD formation in the gas-rich environments of the early Universe.
\end{itemize}

Our results introduce a physically motivated alternative to the classical tidal stripping scenario. We conclude that the cluster environment does not merely strip infalling dwarf galaxies but can actively reshape their structural identity through gas compression and centralized star formation, bridging the gap between the most diffuse and most compact galaxies.

Ultimately, this transformation pathway challenges the static structural classification of extreme dwarf galaxies. These results clearly suggest that the structural extremes of the dwarf galaxy population are not necessarily independent endpoints but can be directly linked through environmentally driven transformation. In this picture, the cluster environment does not simply act as a destructive force. Instead, it is an active driver of galaxy evolution capable of simultaneously dissolving and rebuilding structures. From a theoretical perspective, a natural next step is to quantify how frequently starburst-driven compact system formation operates in clusters and whether it extends to progenitors that do not formally satisfy a UDG structural criterion (or high metallicity outliers, as suggested by a similar scenario). Future observational constraints on stellar populations, particularly age distributions and chemical enrichment patterns, with next-generation spectroscopic facilities, will be required to test this scenario and determine the relative importance of this pathway within the overall CD population. Constraining this transformation observationally will be essential for establishing how frequently such pathways operate and for refining our understanding of environmentally driven galaxy evolution.

\begin{acknowledgements}
 The authors thank the IllustrisTNG team for making their simulations publicly available. The python packages \texttt{matplotlib} \citep{Hunter2007}, \texttt{seaborn} \citep{Waskom2021}, \texttt{numpy} \citep{Harris2020}, \texttt{scipy} \citep{Virtanen2020}, \texttt{pandas} \citep{McKinney2010}, and \texttt{pynbody} \citep{pynbody} were used in parts of this analysis. This research was supported by the Ministry of Science, Technological Development, and Innovation of the Republic of Serbia (MSTDIRS) through contract no. 451-03-33/2026-03/200104, made with the Faculty of Mathematics, University of Belgrade, and contract no. 451-03-33/2026-03/200002, made with the Astronomical Observatory (Belgrade, Serbia).
\end{acknowledgements}

\bibliographystyle{aa}
\bibliography{aa60545-26}

@ARTICLE{Pavlov2025A&A...694A.312P,
       author = {{Pavlov}, Nata{\v{s}}a and {Mitra{\v{s}}inovi{\'c}}, Ana},
        title = "{Puzzling Ultra-Diffuse Galaxy Evolution (PUDGE): I. The existence of a Nube-like galaxy in IllustrisTNG}",
      journal = {\aap},
         year = 2025,
        month = feb,
       volume = {694},
          eid = {A312},
        pages = {A312},
          doi = {10.1051/0004-6361/202452777},
archivePrefix = {arXiv},
       eprint = {2502.04833},
 primaryClass = {astro-ph.GA},
       adsurl = {https://ui.adsabs.harvard.edu/abs/2025A&A...694A.312P}
}

@Article{Hunter2007,
  Author    = {Hunter, J. D.},
  Title     = {Matplotlib: A 2D graphics environment},
  Journal   = {Computing in Science \& Engineering},
  Volume    = {9},
  Number    = {3},
  Pages     = {90--95},
  publisher = {IEEE COMPUTER SOC},
  doi       = {10.1109/MCSE.2007.55},
  year      = 2007
}

@Article{Harris2020,
title         = {Array programming with {NumPy}},
 author        = {Charles R. Harris and K. Jarrod Millman and St{\'{e}}fan J.
                 van der Walt and Ralf Gommers and Pauli Virtanen and David
                 Cournapeau and Eric Wieser and Julian Taylor and Sebastian
                 Berg and Nathaniel J. Smith and Robert Kern and Matti Picus
                 and Stephan Hoyer and Marten H. van Kerkwijk and Matthew
                 Brett and Allan Haldane and Jaime Fern{\'{a}}ndez del
                 R{\'{i}}o and Mark Wiebe and Pearu Peterson and Pierre
                 G{\'{e}}rard-Marchant and Kevin Sheppard and Tyler Reddy and
                 Warren Weckesser and Hameer Abbasi and Christoph Gohlke and
                 Travis E. Oliphant},
 year          = {2020},
 month         = sep,
 journal       = {\nat},
 volume        = {585},
 number        = {7825},
 pages         = {357--362},
 doi           = {10.1038/s41586-020-2649-2},
 publisher     = {Springer Science and Business Media {LLC}},
 url           = {https://doi.org/10.1038/s41586-020-2649-2}
}

@InProceedings{McKinney2010,
  author    = { {W}es {M}c{K}inney },
  title     = { {D}ata {S}tructures for {S}tatistical {C}omputing in {P}ython },
  booktitle = { {P}roceedings of the 9th {P}ython in {S}cience {C}onference },
  pages     = { 56 - 61 },
  year      = { 2010 },
  editor    = { {S}t\'efan van der {W}alt and {J}arrod {M}illman },
  doi       = { 10.25080/Majora-92bf1922-00a }
}

@article{Waskom2021,
    doi = {10.21105/joss.03021},
    url = {https://doi.org/10.21105/joss.03021},
    year = {2021},
    publisher = {The Open Journal},
    volume = {6},
    number = {60},
    pages = {3021},
    author = {Michael L. Waskom},
    title = {seaborn: statistical data visualization},
    journal = {Journal of Open Source Software}
 }

@article{Virtanen2020,
  author  = {Virtanen, Pauli and Gommers, Ralf and Oliphant, Travis E. and
            Haberland, Matt and Reddy, Tyler and Cournapeau, David and
            Burovski, Evgeni and Peterson, Pearu and Weckesser, Warren and
            Bright, Jonathan and {van der Walt}, St{\'e}fan J. and
            Brett, Matthew and Wilson, Joshua and Millman, K. Jarrod and
            Mayorov, Nikolay and Nelson, Andrew R. J. and Jones, Eric and
            Kern, Robert and Larson, Eric and Carey, C J and
            Polat, {\.I}lhan and Feng, Yu and Moore, Eric W. and
            {VanderPlas}, Jake and Laxalde, Denis and Perktold, Josef and
            Cimrman, Robert and Henriksen, Ian and Quintero, E. A. and
            Harris, Charles R. and Archibald, Anne M. and
            Ribeiro, Ant{\^o}nio H. and Pedregosa, Fabian and
            {van Mulbregt}, Paul and {SciPy 1.0 Contributors}},
  title   = {{{SciPy} 1.0: Fundamental Algorithms for Scientific
            Computing in Python}},
  journal = {Nature Methods},
  year    = {2020},
  volume  = {17},
  pages   = {261--272},
  adsurl  = {https://rdcu.be/b08Wh},
  doi     = {10.1038/s41592-019-0686-2}
}

@misc{pynbody,
  author = {{Pontzen}, A. and {Ro{\v s}kar}, R. and {Stinson}, G.~S. and {Woods},
     R. and {Reed}, D.~M. and {Coles}, J. and {Quinn}, T.~R.},
  title = "{pynbody: Astrophysics Simulation Analysis for Python}",
  note = {Astrophysics Source Code Library, ascl:1305.002},
  year = 2013
}

@ARTICLE{2025PASA...42...66M,
       author = {{Mitra{\v{s}}inovi{\'c}}, Ana and {Vukoti{\'c}}, Branislav and {{\v{Z}}i{\v{z}}ak}, Teodora and {Micic}, Miroslav and {{\'C}irkovi{\'c}}, Milan M.},
        title = "{Revisiting the bimodality of galactic habitability in IllustrisTNG}",
      journal = {\pasa},
         year = 2025,
        month = may,
       volume = {42},
          eid = {e066},
        pages = {e066},
          doi = {10.1017/pasa.2025.10040},
archivePrefix = {arXiv},
       eprint = {2505.11048},
 primaryClass = {astro-ph.GA},
       adsurl = {https://ui.adsabs.harvard.edu/abs/2025PASA...42...66M}
}

@ARTICLE{Benavides+2023MNRAS.522.1033B,
       author = {{Benavides}, Jos{\'e} A. and {Sales}, Laura V. and {Abadi}, Mario G. and {Marinacci}, Federico and {Vogelsberger}, Mark and {Hernquist}, Lars},
        title = "{Origin and evolution of ultradiffuse galaxies in different environments}",
      journal = {\mnras},
         year = 2023,
        month = jun,
       volume = {522},
       number = {1},
        pages = {1033-1048},
          doi = {10.1093/mnras/stad1053},
archivePrefix = {arXiv},
       eprint = {2209.07539},
 primaryClass = {astro-ph.GA},
       adsurl = {https://ui.adsabs.harvard.edu/abs/2023MNRAS.522.1033B}
}

@ARTICLE{Bose+2023MNRAS.524.2579B,
       author = {{Bose}, Sownak and {Hadzhiyska}, Boryana and {Barrera}, Monica and {Delgado}, Ana Maria and {Ferlito}, Fulvio and {Frenk}, Carlos and {Hern{\'a}ndez-Aguayo}, C{\'e}sar and {Hernquist}, Lars and {Kannan}, Rahul and {Pakmor}, R{\"u}diger and {Springel}, Volker and {White}, Simon D.~M.},
        title = "{The MillenniumTNG Project: the large-scale clustering of galaxies}",
      journal = {\mnras},
         year = 2023,
        month = sep,
       volume = {524},
       number = {2},
        pages = {2579-2593},
          doi = {10.1093/mnras/stad1097},
archivePrefix = {arXiv},
       eprint = {2210.10065},
 primaryClass = {astro-ph.CO},
       adsurl = {https://ui.adsabs.harvard.edu/abs/2023MNRAS.524.2579B}
}

@ARTICLE{Pakmor+2023MNRAS.524.2539P,
       author = {{Pakmor}, R{\"u}diger and {Springel}, Volker and {Coles}, Jonathan P. and {Guillet}, Thomas and {Pfrommer}, Christoph and {Bose}, Sownak and {Barrera}, Monica and {Delgado}, Ana Maria and {Ferlito}, Fulvio and {Frenk}, Carlos and {Hadzhiyska}, Boryana and {Hern{\'a}ndez-Aguayo}, C{\'e}sar and {Hernquist}, Lars and {Kannan}, Rahul and {White}, Simon D.~M.},
        title = "{The MillenniumTNG Project: the hydrodynamical full physics simulation and a first look at its galaxy clusters}",
      journal = {\mnras},
         year = 2023,
        month = sep,
       volume = {524},
       number = {2},
        pages = {2539-2555},
          doi = {10.1093/mnras/stac3620},
archivePrefix = {arXiv},
       eprint = {2210.10060},
 primaryClass = {astro-ph.CO},
       adsurl = {https://ui.adsabs.harvard.edu/abs/2023MNRAS.524.2539P}
}

@ARTICLE{Nelson+2024-tngcluster,
       author = {{Nelson}, Dylan and {Pillepich}, Annalisa and {Ayromlou}, Mohammadreza and {Lee}, Wonki and {Lehle}, Katrin and {Rohr}, Eric and {Truong}, Nhut},
        title = "{Introducing the TNG-Cluster simulation: Overview and the physical properties of the gaseous intracluster medium}",
      journal = {\aap},
         year = 2024,
        month = jun,
       volume = {686},
          eid = {A157},
        pages = {A157},
          doi = {10.1051/0004-6361/202348608},
archivePrefix = {arXiv},
       eprint = {2311.06338},
 primaryClass = {astro-ph.GA},
       adsurl = {https://ui.adsabs.harvard.edu/abs/2024A&A...686A.157N}
}

@ARTICLE{Rohr+2024A&A...686A..86R,
       author = {{Rohr}, Eric and {Pillepich}, Annalisa and {Nelson}, Dylan and {Ayromlou}, Mohammadreza and {Zinger}, Elad},
        title = "{The hot circumgalactic media of massive cluster satellites in the TNG-Cluster simulation: Existence and detectability}",
      journal = {\aap},
         year = 2024,
        month = jun,
       volume = {686},
          eid = {A86},
        pages = {A86},
          doi = {10.1051/0004-6361/202348583},
archivePrefix = {arXiv},
       eprint = {2311.06337},
 primaryClass = {astro-ph.GA},
       adsurl = {https://ui.adsabs.harvard.edu/abs/2024A&A...686A..86R}
}

@ARTICLE{Lehle+2024A&A...687A.129L,
       author = {{Lehle}, Katrin and {Nelson}, Dylan and {Pillepich}, Annalisa and {Truong}, Nhut and {Rohr}, Eric},
        title = "{The heart of galaxy clusters: Demographics and physical properties of cool-core and non-cool-core halos in the TNG-Cluster simulation}",
      journal = {\aap},
         year = 2024,
        month = jul,
       volume = {687},
          eid = {A129},
        pages = {A129},
          doi = {10.1051/0004-6361/202348609},
archivePrefix = {arXiv},
       eprint = {2311.06333},
 primaryClass = {astro-ph.GA},
       adsurl = {https://ui.adsabs.harvard.edu/abs/2024A&A...687A.129L}
}

@ARTICLE{Ayromlou+2024A&A...690A..20A,
       author = {{Ayromlou}, Mohammadreza and {Nelson}, Dylan and {Pillepich}, Annalisa and {Rohr}, Eric and {Truong}, Nhut and {Li}, Yuan and {Simionescu}, Aurora and {Lehle}, Katrin and {Lee}, Wonki},
        title = "{An atlas of gas motions in the TNG-Cluster simulation: From cluster cores to the outskirts}",
      journal = {\aap},
         year = 2024,
        month = oct,
       volume = {690},
          eid = {A20},
        pages = {A20},
          doi = {10.1051/0004-6361/202348612},
archivePrefix = {arXiv},
       eprint = {2311.06339},
 primaryClass = {astro-ph.GA},
       adsurl = {https://ui.adsabs.harvard.edu/abs/2024A&A...690A..20A}
}

@ARTICLE{Kravtsov+Borgani2012ARA&A..50..353K,
       author = {{Kravtsov}, Andrey V. and {Borgani}, Stefano},
        title = "{Formation of Galaxy Clusters}",
      journal = {\araa},
         year = 2012,
        month = sep,
       volume = {50},
        pages = {353-409},
          doi = {10.1146/annurev-astro-081811-125502},
archivePrefix = {arXiv},
       eprint = {1205.5556},
 primaryClass = {astro-ph.CO},
       adsurl = {https://ui.adsabs.harvard.edu/abs/2012ARA&A..50..353K}
}

@ARTICLE{Voit2005RvMP...77..207V,
       author = {{Voit}, G. Mark},
        title = "{Tracing cosmic evolution with clusters of galaxies}",
      journal = {Reviews of Modern Physics},
         year = 2005,
        month = apr,
       volume = {77},
       number = {1},
        pages = {207-258},
          doi = {10.1103/RevModPhys.77.207},
archivePrefix = {arXiv},
       eprint = {astro-ph/0410173},
 primaryClass = {astro-ph},
       adsurl = {https://ui.adsabs.harvard.edu/abs/2005RvMP...77..207V}
}

@ARTICLE{Matthews+1964,
       author = {{Matthews}, Thomas A. and {Morgan}, William W. and {Schmidt}, Maarten},
        title = "{A Discussion of Galaxies Indentified with Radio Sources.}",
      journal = {\apj},
         year = 1964,
        month = jul,
       volume = {140},
        pages = {35},
          doi = {10.1086/147890},
       adsurl = {https://ui.adsabs.harvard.edu/abs/1964ApJ...140...35M}
}

@ARTICLE{tormen1998,
       author = {{Tormen}, Giuseppe and {Diaferio}, Antonaldo and {Syer}, D.},
        title = "{Survival of substructure within dark matter haloes}",
      journal = {\mnras},
         year = 1998,
        month = sep,
       volume = {299},
       number = {3},
        pages = {728-742},
          doi = {10.1046/j.1365-8711.1998.01775.x},
archivePrefix = {arXiv},
       eprint = {astro-ph/9712222},
 primaryClass = {astro-ph},
       adsurl = {https://ui.adsabs.harvard.edu/abs/1998MNRAS.299..728T}
}

@ARTICLE{Knebe2004PASA,
       author = {{Knebe}, Alexander and {Gill}, Stuart P.~D. and {Gibson}, Brad K.},
        title = "{Interactions of Satellite Galaxies in Cosmological Dark Matter Halos}",
      journal = {\pasa},
         year = 2004,
        month = jan,
       volume = {21},
       number = {2},
        pages = {216-221},
          doi = {10.1071/AS04018},
archivePrefix = {arXiv},
       eprint = {astro-ph/0402390},
 primaryClass = {astro-ph},
       adsurl = {https://ui.adsabs.harvard.edu/abs/2004PASA...21..216K}
}

@ARTICLE{tt1972,
       author = {{Toomre}, Alar and {Toomre}, Juri},
        title = "{Galactic Bridges and Tails}",
      journal = {\apj},
         year = 1972,
        month = dec,
       volume = {178},
        pages = {623-666},
          doi = {10.1086/151823},
       adsurl = {https://ui.adsabs.harvard.edu/abs/1972ApJ...178..623T}
}

@ARTICLE{barnes&hernquist1992,
       author = {{Barnes}, Joshua E. and {Hernquist}, Lars},
        title = "{Dynamics of interacting galaxies.}",
      journal = {\araa},
         year = 1992,
        month = jan,
       volume = {30},
        pages = {705-742},
          doi = {10.1146/annurev.aa.30.090192.003421},
       adsurl = {https://ui.adsabs.harvard.edu/abs/1992ARA&A..30..705B}
}

@ARTICLE{Eneev1973,
       author = {{Eneev}, T.~M. and {Kozlov}, N.~N. and {Sunyaev}, R.~A.},
        title = "{Tidal Interaction of Galaxies}",
      journal = {\aap},
         year = 1973,
        month = jan,
       volume = {22},
        pages = {41},
       adsurl = {https://ui.adsabs.harvard.edu/abs/1973A&A....22...41E}
}

@ARTICLE{Tutukov2006,
       author = {{Tutukov}, A.~V. and {Fedorova}, A.~V.},
        title = "{The role of close passages of galaxies and the asymmetry of their dark haloes in the formation of their spiral patterns}",
      journal = {Astronomy Reports},
         year = 2006,
        month = oct,
       volume = {50},
       number = {10},
        pages = {785-801},
          doi = {10.1134/S1063772906100039},
       adsurl = {https://ui.adsabs.harvard.edu/abs/2006ARep...50..785T}
}

@ARTICLE{Dubinski&Chakrabarty2009,
       author = {{Dubinski}, John and {Chakrabarty}, Dalia},
        title = "{Warps and Bars from the External Tidal Torques of Tumbling Dark Halos}",
      journal = {\apj},
         year = 2009,
        month = oct,
       volume = {703},
       number = {2},
        pages = {2068-2081},
          doi = {10.1088/0004-637X/703/2/2068},
archivePrefix = {arXiv},
       eprint = {0908.0168},
 primaryClass = {astro-ph.GA},
       adsurl = {https://ui.adsabs.harvard.edu/abs/2009ApJ...703.2068D}
}

@ARTICLE{Mitrasinovic+Micic2023,
       author = {{Mitra{\v{s}}inovi{\'c}}, A. and {Micic}, M.},
        title = "{The role of impact parameter in typical close galaxy flybys}",
      journal = {\pasa},
         year = 2023,
        month = may,
       volume = {40},
          eid = {e024},
        pages = {e024},
          doi = {10.1017/pasa.2023.23},
archivePrefix = {arXiv},
       eprint = {2304.07751},
 primaryClass = {astro-ph.GA},
       adsurl = {https://ui.adsabs.harvard.edu/abs/2023PASA...40...24M}
}

@ARTICLE{Pettitt+Wadsley2018,
       author = {{Pettitt}, Alex R. and {Wadsley}, J.~W.},
        title = "{Bars and spirals in tidal interactions with an ensemble of galaxy mass models}",
      journal = {\mnras},
         year = 2018,
        month = mar,
       volume = {474},
       number = {4},
        pages = {5645-5671},
          doi = {10.1093/mnras/stx3129},
archivePrefix = {arXiv},
       eprint = {1712.00882},
 primaryClass = {astro-ph.GA},
       adsurl = {https://ui.adsabs.harvard.edu/abs/2018MNRAS.474.5645P}
}

@ARTICLE{sinha2012,
       author = {{Sinha}, Manodeep and {Holley-Bockelmann}, Kelly},
        title = "{A First Look at Galaxy Flyby Interactions. I. Characterizing the Frequency of Flybys in a Cosmological Context}",
      journal = {\apj},
         year = 2012,
        month = may,
       volume = {751},
       number = {1},
          eid = {17},
        pages = {17},
          doi = {10.1088/0004-637X/751/1/17},
archivePrefix = {arXiv},
       eprint = {1103.1675},
 primaryClass = {astro-ph.CO},
       adsurl = {https://ui.adsabs.harvard.edu/abs/2012ApJ...751...17S}
}

@ARTICLE{shan2019,
       author = {{An}, Sung-Ho and {Kim}, Juhan and {Moon}, Jun-Sung and {Yoon}, Suk-Jin},
        title = "{Living with Neighbors. II. Statistical Analysis of Flybys and Mergers of Dark Matter Halos in Cosmological Simulations}",
      journal = {\apj},
         year = 2019,
        month = dec,
       volume = {887},
       number = {1},
          eid = {59},
        pages = {59},
          doi = {10.3847/1538-4357/ab535f},
archivePrefix = {arXiv},
       eprint = {1911.11782},
 primaryClass = {astro-ph.GA},
       adsurl = {https://ui.adsabs.harvard.edu/abs/2019ApJ...887...59A}
}

@ARTICLE{moore1996,
       author = {{Moore}, Ben and {Katz}, Neal and {Lake}, George and {Dressler}, Alan and {Oemler}, Augustus},
        title = "{Galaxy harassment and the evolution of clusters of galaxies}",
      journal = {\nat},
         year = 1996,
        month = feb,
       volume = {379},
       number = {6566},
        pages = {613-616},
          doi = {10.1038/379613a0},
archivePrefix = {arXiv},
       eprint = {astro-ph/9510034},
 primaryClass = {astro-ph},
       adsurl = {https://ui.adsabs.harvard.edu/abs/1996Natur.379..613M}
}

@ARTICLE{moore1998,
       author = {{Moore}, Ben and {Lake}, George and {Katz}, Neal},
        title = "{Morphological Transformation from Galaxy Harassment}",
      journal = {\apj},
         year = 1998,
        month = mar,
       volume = {495},
       number = {1},
        pages = {139-151},
          doi = {10.1086/305264},
archivePrefix = {arXiv},
       eprint = {astro-ph/9701211},
 primaryClass = {astro-ph},
       adsurl = {https://ui.adsabs.harvard.edu/abs/1998ApJ...495..139M}
}

@ARTICLE{moore1999,
       author = {{Moore}, Ben and {Lake}, George and {Quinn}, Thomas and {Stadel}, Joachim},
        title = "{On the survival and destruction of spiral galaxies in clusters}",
      journal = {\mnras},
         year = 1999,
        month = apr,
       volume = {304},
       number = {3},
        pages = {465-474},
          doi = {10.1046/j.1365-8711.1999.02345.x},
archivePrefix = {arXiv},
       eprint = {astro-ph/9811127},
 primaryClass = {astro-ph},
       adsurl = {https://ui.adsabs.harvard.edu/abs/1999MNRAS.304..465M}
}

@ARTICLE{mastropietro2005,
       author = {{Mastropietro}, Chiara and {Moore}, Ben and {Mayer}, Lucio and {Debattista}, Victor P. and {Piffaretti}, Rocco and {Stadel}, Joachim},
        title = "{Morphological evolution of discs in clusters}",
      journal = {\mnras},
         year = 2005,
        month = dec,
       volume = {364},
       number = {2},
        pages = {607-619},
          doi = {10.1111/j.1365-2966.2005.09579.x},
archivePrefix = {arXiv},
       eprint = {astro-ph/0411648},
 primaryClass = {astro-ph},
       adsurl = {https://ui.adsabs.harvard.edu/abs/2005MNRAS.364..607M}
}

@ARTICLE{Lin2023,
       author = {{Lin}, Xuchen and {Wang}, Jing and {Kilborn}, Virginia and {Peng}, Eric W. and {Cortese}, Luca and {Boselli}, Alessandro and {Liang}, Ze-Zhong and {Lee}, Bumhyun and {Yang}, Dong and {Catinella}, Barbara and {Deg}, N. and {D{\'e}nes}, H. and {Elagali}, Ahmed and {Kamphuis}, P. and {Koribalski}, B.~S. and {Lee-Waddell}, K. and {Rhee}, Jonghwan and {Shao}, Li and {Spekkens}, Kristine and {Staveley-Smith}, Lister and {Westmeier}, T. and {Wong}, O. Ivy and {Bekki}, Kenji and {Bosma}, Albert and {Du}, Min and {Ho}, Luis C. and {Madrid}, Juan P. and {Verdes-Montenegro}, Lourdes and {Wang}, Huiyuan and {Wang}, Shun},
        title = "{FAST-ASKAP Synergy: Quantifying Coexistent Tidal and Ram Pressure Strippings in the NGC 4636 Group}",
      journal = {\apj},
         year = 2023,
        month = oct,
       volume = {956},
       number = {2},
          eid = {148},
        pages = {148},
          doi = {10.3847/1538-4357/accea2},
archivePrefix = {arXiv},
       eprint = {2304.09795},
 primaryClass = {astro-ph.GA},
       adsurl = {https://ui.adsabs.harvard.edu/abs/2023ApJ...956..148L}
}

@ARTICLE{Reynolds2022,
       author = {{Reynolds}, T.~N. and {Catinella}, B. and {Cortese}, L. and {Westmeier}, T. and {Meurer}, G.~R. and {Shao}, L. and {Obreschkow}, D. and {Rom{\'a}n}, J. and {Verdes-Montenegro}, L. and {Deg}, N. and {D{\'e}nes}, H. and {For}, B. -Q. and {Kleiner}, D. and {Koribalski}, B.~S. and {Lee-Waddell}, K. and {Murugeshan}, C. and {Oh}, S. -H. and {Rhee}, J. and {Spekkens}, K. and {Staveley-Smith}, L. and {Stevens}, A.~R.~H. and {van der Hulst}, J.~M. and {Wang}, J. and {Wong}, O.~I. and {Holwerda}, B.~W. and {Bosma}, A. and {Madrid}, J.~P. and {Bekki}, K.},
        title = "{WALLABY pilot survey: H I gas disc truncation and star formation of galaxies falling into the Hydra I cluster}",
      journal = {\mnras},
         year = 2022,
        month = feb,
       volume = {510},
       number = {2},
        pages = {1716-1732},
          doi = {10.1093/mnras/stab3522},
archivePrefix = {arXiv},
       eprint = {2112.00231},
 primaryClass = {astro-ph.GA},
       adsurl = {https://ui.adsabs.harvard.edu/abs/2022MNRAS.510.1716R}
}

@ARTICLE{Holwerda2023,
       author = {{Holwerda}, Benne~W. and {Bigiel}, Frank and {Bosma}, Albert and {Courtois}, Helene M. and {Deg}, Nathan and {D{\'e}nes}, Helga and {Elagali}, Ahmed and {For}, Bi-Qing and {Koribalski}, Baerbel and {Leahy}, Denis A. and {Lee-Waddell}, Karen and {L{\'o}pez-S{\'a}nchez}, {\'A}ngel R. and {Oh}, Se-Heon and {Reynolds}, Tristan N. and {Rhee}, Jonghwan and {Spekkens}, Kristine and {Wang}, Jing and {Westmeier}, Tobias and {Wong}, O. Ivy},
        title = "{WALLABY Pilot Survey: hydra cluster galaxies UV and H I morphometrics}",
      journal = {\mnras},
         year = 2023,
        month = may,
       volume = {521},
       number = {1},
        pages = {1502-1517},
          doi = {10.1093/mnras/stad602},
archivePrefix = {arXiv},
       eprint = {2302.07963},
 primaryClass = {astro-ph.GA},
       adsurl = {https://ui.adsabs.harvard.edu/abs/2023MNRAS.521.1502H}
}

@ARTICLE{Holwerda2025,
       author = {{Holwerda}, Benne~W. and {D{\'e}nes}, Helga and {Rhee}, Jonghwan and {Leahy}, Denis and {Koribalski}, B{\"a}rbel Silvia and {Yu}, Niankun and {Deg}, Nathan and {Westmeier}, T. and {Lee-Waddell}, Karen and {Ascasibar}, Yago and {Saraf}, Manasvee and {Lin}, Xuchen and {Catinella}, Barbara and {Hess}, Kelley},
        title = "{WALLABY Pilot Survey: kNN identification of perturbed galaxies through H 1 morphometrics}",
      journal = {\pasa},
         year = 2025,
        month = jan,
       volume = {42},
          eid = {e028},
        pages = {e028},
          doi = {10.1017/pasa.2025.5},
archivePrefix = {arXiv},
       eprint = {2501.10563},
 primaryClass = {astro-ph.GA},
       adsurl = {https://ui.adsabs.harvard.edu/abs/2025PASA...42...28H}
}

@ARTICLE{Lopes+2024MNRAS.527L..19L,
       author = {{Lopes}, Paulo A.~A. and {Ribeiro}, Andr{\'e} L.~B. and {Brambila}, Douglas},
        title = "{The role of groups in galaxy evolution: compelling evidence of pre-processing out to the turnaround radius of clusters}",
      journal = {\mnras},
         year = 2024,
        month = jan,
       volume = {527},
       number = {1},
        pages = {L19-L25},
          doi = {10.1093/mnrasl/slad134},
archivePrefix = {arXiv},
       eprint = {2309.11578},
 primaryClass = {astro-ph.CO},
       adsurl = {https://ui.adsabs.harvard.edu/abs/2024MNRAS.527L..19L}
}

@ARTICLE{Piraino-Cerda+2024MNRAS.528..919P,
       author = {{Piraino-Cerda}, Franco and {Jaff{\'e}}, Yara L. and {Louren{\c{c}}o}, Ana C. and {Crossett}, Jacob P. and {Salinas}, Vicente and {Kim}, Duho and {Sheen}, Yun-Kyeong and {Kelkar}, Kshitija and {Pallero}, Diego and {Bravo-Alfaro}, Hector},
        title = "{Pre- and post-processing of cluster galaxies out to 5 {\texttimes} R$_{200}$: the extreme case of A2670}",
      journal = {\mnras},
         year = 2024,
        month = feb,
       volume = {528},
       number = {1},
        pages = {919-936},
          doi = {10.1093/mnras/stad3957},
archivePrefix = {arXiv},
       eprint = {2401.06973},
 primaryClass = {astro-ph.GA},
       adsurl = {https://ui.adsabs.harvard.edu/abs/2024MNRAS.528..919P}
}

@ARTICLE{Xie+2025A&A...698A..73X,
       author = {{Xie}, Lizhi and {De Lucia}, Gabriella and {Fossati}, Matteo and {Fontanot}, Fabio and {Hirschmann}, Michaela},
        title = "{The impact of ram pressure on cluster galaxies, insights from GAEA and TNG}",
      journal = {\aap},
         year = 2025,
        month = jun,
       volume = {698},
          eid = {A73},
        pages = {A73},
          doi = {10.1051/0004-6361/202553915},
archivePrefix = {arXiv},
       eprint = {2504.12863},
 primaryClass = {astro-ph.GA},
       adsurl = {https://ui.adsabs.harvard.edu/abs/2025A&A...698A..73X}
}

@ARTICLE{Paul+2017,
       author = {{Paul}, S. and {John}, R.~S. and {Gupta}, P. and {Kumar}, H.},
        title = "{Understanding `galaxy groups' as a unique structure in the universe}",
      journal = {\mnras},
         year = 2017,
        month = oct,
       volume = {471},
       number = {1},
        pages = {2-11},
          doi = {10.1093/mnras/stx1488},
archivePrefix = {arXiv},
       eprint = {1706.01916},
 primaryClass = {astro-ph.CO},
       adsurl = {https://ui.adsabs.harvard.edu/abs/2017MNRAS.471....2P}
}

@ARTICLE{Mitrasinovic2025A&A...703L..16M,
       author = {{Mitra{\v{s}}inovi{\'c}}, Ana},
        title = "{Living on the edge: A quantitative warning on boundary artifacts in the IllustrisTNG}",
      journal = {\aap},
         year = 2025,
        month = nov,
       volume = {703},
          eid = {L16},
        pages = {L16},
          doi = {10.1051/0004-6361/202557557},
archivePrefix = {arXiv},
       eprint = {2511.03562},
 primaryClass = {astro-ph.CO},
       adsurl = {https://ui.adsabs.harvard.edu/abs/2025A&A...703L..16M}
}

@ARTICLE{TNGmethods2017,
       author = {{Weinberger}, Rainer and {Springel}, Volker and {Hernquist}, Lars and {Pillepich}, Annalisa and {Marinacci}, Federico and {Pakmor}, R{\"u}diger and {Nelson}, Dylan and {Genel}, Shy and {Vogelsberger}, Mark and {Naiman}, Jill and {Torrey}, Paul},
        title = "{Simulating galaxy formation with black hole driven thermal and kinetic feedback}",
      journal = {\mnras},
         year = 2017,
        month = mar,
       volume = {465},
       number = {3},
        pages = {3291-3308},
          doi = {10.1093/mnras/stw2944},
archivePrefix = {arXiv},
       eprint = {1607.03486},
 primaryClass = {astro-ph.GA},
       adsurl = {https://ui.adsabs.harvard.edu/abs/2017MNRAS.465.3291W}
}

@ARTICLE{TNGmethods2018,
       author = {{Pillepich}, Annalisa and {Springel}, Volker and {Nelson}, Dylan and {Genel}, Shy and {Naiman}, Jill and {Pakmor}, R{\"u}diger and {Hernquist}, Lars and {Torrey}, Paul and {Vogelsberger}, Mark and {Weinberger}, Rainer and {Marinacci}, Federico},
        title = "{Simulating galaxy formation with the IllustrisTNG model}",
      journal = {\mnras},
         year = 2018,
        month = jan,
       volume = {473},
       number = {3},
        pages = {4077-4106},
          doi = {10.1093/mnras/stx2656},
archivePrefix = {arXiv},
       eprint = {1703.02970},
 primaryClass = {astro-ph.GA},
       adsurl = {https://ui.adsabs.harvard.edu/abs/2018MNRAS.473.4077P}
}

@ARTICLE{Marinacci+2018,
       author = {{Marinacci}, Federico and {Vogelsberger}, Mark and {Pakmor}, R{\"u}diger and {Torrey}, Paul and {Springel}, Volker and {Hernquist}, Lars and {Nelson}, Dylan and {Weinberger}, Rainer and {Pillepich}, Annalisa and {Naiman}, Jill and {Genel}, Shy},
        title = "{First results from the IllustrisTNG simulations: radio haloes and magnetic fields}",
      journal = {\mnras},
         year = 2018,
        month = nov,
       volume = {480},
       number = {4},
        pages = {5113-5139},
          doi = {10.1093/mnras/sty2206},
archivePrefix = {arXiv},
       eprint = {1707.03396},
 primaryClass = {astro-ph.CO},
       adsurl = {https://ui.adsabs.harvard.edu/abs/2018MNRAS.480.5113M}
}

@ARTICLE{Pillepich+2018,
       author = {{Pillepich}, Annalisa and {Nelson}, Dylan and {Hernquist}, Lars and {Springel}, Volker and {Pakmor}, R{\"u}diger and {Torrey}, Paul and {Weinberger}, Rainer and {Genel}, Shy and {Naiman}, Jill P. and {Marinacci}, Federico and {Vogelsberger}, Mark},
        title = "{First results from the IllustrisTNG simulations: the stellar mass content of groups and clusters of galaxies}",
      journal = {\mnras},
         year = 2018,
        month = mar,
       volume = {475},
       number = {1},
        pages = {648-675},
          doi = {10.1093/mnras/stx3112},
archivePrefix = {arXiv},
       eprint = {1707.03406},
 primaryClass = {astro-ph.GA},
       adsurl = {https://ui.adsabs.harvard.edu/abs/2018MNRAS.475..648P}
}

@ARTICLE{Naiman+2018,
       author = {{Naiman}, Jill P. and {Pillepich}, Annalisa and {Springel}, Volker and {Ramirez-Ruiz}, Enrico and {Torrey}, Paul and {Vogelsberger}, Mark and {Pakmor}, R{\"u}diger and {Nelson}, Dylan and {Marinacci}, Federico and {Hernquist}, Lars and {Weinberger}, Rainer and {Genel}, Shy},
        title = "{First results from the IllustrisTNG simulations: a tale of two elements - chemical evolution of magnesium and europium}",
      journal = {\mnras},
         year = 2018,
        month = jun,
       volume = {477},
       number = {1},
        pages = {1206-1224},
          doi = {10.1093/mnras/sty618},
archivePrefix = {arXiv},
       eprint = {1707.03401},
 primaryClass = {astro-ph.GA},
       adsurl = {https://ui.adsabs.harvard.edu/abs/2018MNRAS.477.1206N}
}

@ARTICLE{Nelson+2018,
       author = {{Nelson}, Dylan and {Pillepich}, Annalisa and {Springel}, Volker and {Weinberger}, Rainer and {Hernquist}, Lars and {Pakmor}, R{\"u}diger and {Genel}, Shy and {Torrey}, Paul and {Vogelsberger}, Mark and {Kauffmann}, Guinevere and {Marinacci}, Federico and {Naiman}, Jill},
        title = "{First results from the IllustrisTNG simulations: the galaxy colour bimodality}",
      journal = {\mnras},
         year = 2018,
        month = mar,
       volume = {475},
       number = {1},
        pages = {624-647},
          doi = {10.1093/mnras/stx3040},
archivePrefix = {arXiv},
       eprint = {1707.03395},
 primaryClass = {astro-ph.GA},
       adsurl = {https://ui.adsabs.harvard.edu/abs/2018MNRAS.475..624N}
}

@ARTICLE{Springel+2018,
       author = {{Springel}, Volker and {Pakmor}, R{\"u}diger and {Pillepich}, Annalisa and {Weinberger}, Rainer and {Nelson}, Dylan and {Hernquist}, Lars and {Vogelsberger}, Mark and {Genel}, Shy and {Torrey}, Paul and {Marinacci}, Federico and {Naiman}, Jill},
        title = "{First results from the IllustrisTNG simulations: matter and galaxy clustering}",
      journal = {\mnras},
         year = 2018,
        month = mar,
       volume = {475},
       number = {1},
        pages = {676-698},
          doi = {10.1093/mnras/stx3304},
archivePrefix = {arXiv},
       eprint = {1707.03397},
 primaryClass = {astro-ph.GA},
       adsurl = {https://ui.adsabs.harvard.edu/abs/2018MNRAS.475..676S}
}

@ARTICLE{Nelson+2019ComAC,
       author = {{Nelson}, Dylan and {Springel}, Volker and {Pillepich}, Annalisa and {Rodriguez-Gomez}, Vicente and {Torrey}, Paul and {Genel}, Shy and {Vogelsberger}, Mark and {Pakmor}, Ruediger and {Marinacci}, Federico and {Weinberger}, Rainer and {Kelley}, Luke and {Lovell}, Mark and {Diemer}, Benedikt and {Hernquist}, Lars},
        title = "{The IllustrisTNG simulations: public data release}",
      journal = {Computational Astrophysics and Cosmology},
         year = 2019,
        month = may,
       volume = {6},
       number = {1},
          eid = {2},
        pages = {2},
          doi = {10.1186/s40668-019-0028-x},
archivePrefix = {arXiv},
       eprint = {1812.05609},
 primaryClass = {astro-ph.GA},
       adsurl = {https://ui.adsabs.harvard.edu/abs/2019ComAC...6....2N}
}

@ARTICLE{PlanckColab+2016,
       author = {{Planck Collaboration} and {Ade}, P.~A.~R. and {Aghanim}, N. and {Arnaud}, M. and {Ashdown}, M. and {Aumont}, J. and {Baccigalupi}, C. and {Banday}, A.~J. and {Barreiro}, R.~B. and {Bartlett}, J.~G. and {Bartolo}, N. and {Battaner}, E. and {Battye}, R. and {Benabed}, K. and {Beno{\^\i}t}, A. and {Benoit-L{\'e}vy}, A. and {Bernard}, J. -P. and {Bersanelli}, M. and {Bielewicz}, P. and {Bock}, J.~J. and {Bonaldi}, A. and {Bonavera}, L. and {Bond}, J.~R. and {Borrill}, J. and {Bouchet}, F.~R. and {Boulanger}, F. and {Bucher}, M. and {Burigana}, C. and {Butler}, R.~C. and {Calabrese}, E. and {Cardoso}, J. -F. and {Catalano}, A. and {Challinor}, A. and {Chamballu}, A. and {Chary}, R. -R. and {Chiang}, H.~C. and {Chluba}, J. and {Christensen}, P.~R. and {Church}, S. and {Clements}, D.~L. and {Colombi}, S. and {Colombo}, L.~P.~L. and {Combet}, C. and {Coulais}, A. and {Crill}, B.~P. and {Curto}, A. and {Cuttaia}, F. and {Danese}, L. and {Davies}, R.~D. and {Davis}, R.~J. and {de Bernardis}, P. and {de Rosa}, A. and {de Zotti}, G. and {Delabrouille}, J. and {D{\'e}sert}, F. -X. and {Di Valentino}, E. and {Dickinson}, C. and {Diego}, J.~M. and {Dolag}, K. and {Dole}, H. and {Donzelli}, S. and {Dor{\'e}}, O. and {Douspis}, M. and {Ducout}, A. and {Dunkley}, J. and {Dupac}, X. and {Efstathiou}, G. and {Elsner}, F. and {En{\ss}lin}, T.~A. and {Eriksen}, H.~K. and {Farhang}, M. and {Fergusson}, J. and {Finelli}, F. and {Forni}, O. and {Frailis}, M. and {Fraisse}, A.~A. and {Franceschi}, E. and {Frejsel}, A. and {Galeotta}, S. and {Galli}, S. and {Ganga}, K. and {Gauthier}, C. and {Gerbino}, M. and {Ghosh}, T. and {Giard}, M. and {Giraud-H{\'e}raud}, Y. and {Giusarma}, E. and {Gjerl{\o}w}, E. and {Gonz{\'a}lez-Nuevo}, J. and {G{\'o}rski}, K.~M. and {Gratton}, S. and {Gregorio}, A. and {Gruppuso}, A. and {Gudmundsson}, J.~E. and {Hamann}, J. and {Hansen}, F.~K. and {Hanson}, D. and {Harrison}, D.~L. and {Helou}, G. and {Henrot-Versill{\'e}}, S. and {Hern{\'a}ndez-Monteagudo}, C. and {Herranz}, D. and {Hildebrandt}, S.~R. and {Hivon}, E. and {Hobson}, M. and {Holmes}, W.~A. and {Hornstrup}, A. and {Hovest}, W. and {Huang}, Z. and {Huffenberger}, K.~M. and {Hurier}, G. and {Jaffe}, A.~H. and {Jaffe}, T.~R. and {Jones}, W.~C. and {Juvela}, M. and {Keih{\"a}nen}, E. and {Keskitalo}, R. and {Kisner}, T.~S. and {Kneissl}, R. and {Knoche}, J. and {Knox}, L. and {Kunz}, M. and {Kurki-Suonio}, H. and {Lagache}, G. and {L{\"a}hteenm{\"a}ki}, A. and {Lamarre}, J. -M. and {Lasenby}, A. and {Lattanzi}, M. and {Lawrence}, C.~R. and {Leahy}, J.~P. and {Leonardi}, R. and {Lesgourgues}, J. and {Levrier}, F. and {Lewis}, A. and {Liguori}, M. and {Lilje}, P.~B. and {Linden-V{\o}rnle}, M. and {L{\'o}pez-Caniego}, M. and {Lubin}, P.~M. and {Mac{\'\i}as-P{\'e}rez}, J.~F. and {Maggio}, G. and {Maino}, D. and {Mandolesi}, N. and {Mangilli}, A. and {Marchini}, A. and {Maris}, M. and {Martin}, P.~G. and {Martinelli}, M. and {Mart{\'\i}nez-Gonz{\'a}lez}, E. and {Masi}, S. and {Matarrese}, S. and {McGehee}, P. and {Meinhold}, P.~R. and {Melchiorri}, A. and {Melin}, J. -B. and {Mendes}, L. and {Mennella}, A. and {Migliaccio}, M. and {Millea}, M. and {Mitra}, S. and {Miville-Desch{\^e}nes}, M. -A. and {Moneti}, A. and {Montier}, L. and {Morgante}, G. and {Mortlock}, D. and {Moss}, A. and {Munshi}, D. and {Murphy}, J.~A. and {Naselsky}, P. and {Nati}, F. and {Natoli}, P. and {Netterfield}, C.~B. and {N{\o}rgaard-Nielsen}, H.~U. and {Noviello}, F. and {Novikov}, D. and {Novikov}, I. and {Oxborrow}, C.~A. and {Paci}, F. and {Pagano}, L. and {Pajot}, F. and {Paladini}, R. and {Paoletti}, D. and {Partridge}, B. and {Pasian}, F. and {Patanchon}, G. and {Pearson}, T.~J. and {Perdereau}, O. and {Perotto}, L. and {Perrotta}, F. and {Pettorino}, V. and {Piacentini}, F. and {Piat}, M. and {Pierpaoli}, E. and {Pietrobon}, D. and {Plaszczynski}, S. and {Pointecouteau}, E. and {Polenta}, G. and {Popa}, L. and {Pratt}, G.~W. and {Pr{\'e}zeau}, G. and {Prunet}, S. and {Puget}, J. -L. and {Rachen}, J.~P. and {Reach}, W.~T. and {Rebolo}, R. and {Reinecke}, M. and {Remazeilles}, M. and {Renault}, C. and {Renzi}, A. and {Ristorcelli}, I. and {Rocha}, G. and {Rosset}, C. and {Rossetti}, M. and {Roudier}, G. and {Rouill{\'e} d'Orfeuil}, B. and {Rowan-Robinson}, M. and {Rubi{\~n}o-Mart{\'\i}n}, J.~A. and {Rusholme}, B. and {Said}, N. and {Salvatelli}, V. and {Salvati}, L. and {Sandri}, M. and {Santos}, D. and {Savelainen}, M. and {Savini}, G. and {Scott}, D. and {Seiffert}, M.~D. and {Serra}, P. and {Shellard}, E.~P.~S. and {Spencer}, L.~D. and {Spinelli}, M. and {Stolyarov}, V. and {Stompor}, R. and {Sudiwala}, R. and {Sunyaev}, R. and {Sutton}, D. and {Suur-Uski}, A. -S. and {Sygnet}, J. -F. and {Tauber}, J.~A. and {Terenzi}, L. and {Toffolatti}, L. and {Tomasi}, M. and {Tristram}, M. and {Trombetti}, T. and {Tucci}, M. and {Tuovinen}, J. and {T{\"u}rler}, M. and {Umana}, G. and {Valenziano}, L. and {Valiviita}, J. and {Van Tent}, F. and {Vielva}, P. and {Villa}, F. and {Wade}, L.~A. and {Wandelt}, B.~D. and {Wehus}, I.~K. and {White}, M. and {White}, S.~D.~M. and {Wilkinson}, A. and {Yvon}, D. and {Zacchei}, A. and {Zonca}, A.},
        title = "{Planck 2015 results. XIII. Cosmological parameters}",
      journal = {\aap},
         year = 2016,
        month = sep,
       volume = {594},
          eid = {A13},
        pages = {A13},
          doi = {10.1051/0004-6361/201525830},
archivePrefix = {arXiv},
       eprint = {1502.01589},
 primaryClass = {astro-ph.CO},
       adsurl = {https://ui.adsabs.harvard.edu/abs/2016A&A...594A..13P}
}

@ARTICLE{Onions+2012MNRAS.423.1200O,
       author = {{Onions}, Julian and {Knebe}, Alexander and {Pearce}, Frazer R. and {Muldrew}, Stuart I. and {Lux}, Hanni and {Knollmann}, Steffen R. and {Ascasibar}, Yago and {Behroozi}, Peter and {Elahi}, Pascal and {Han}, Jiaxin and {Maciejewski}, Michal and {Merch{\'a}n}, Manuel E. and {Neyrinck}, Mark and {Ruiz}, Andr{\'e}s. N. and {Sgr{\'o}}, Mario A. and {Springel}, Volker and {Tweed}, Dylan},
        title = "{Subhaloes going Notts: the subhalo-finder comparison project}",
      journal = {\mnras},
         year = 2012,
        month = jun,
       volume = {423},
       number = {2},
        pages = {1200-1214},
          doi = {10.1111/j.1365-2966.2012.20947.x},
archivePrefix = {arXiv},
       eprint = {1203.3695},
 primaryClass = {astro-ph.CO},
       adsurl = {https://ui.adsabs.harvard.edu/abs/2012MNRAS.423.1200O}
}

@ARTICLE{Springel2010AREPO,
       author = {{Springel}, Volker},
        title = "{E pur si muove: Galilean-invariant cosmological hydrodynamical simulations on a moving mesh}",
      journal = {\mnras},
         year = 2010,
        month = jan,
       volume = {401},
       number = {2},
        pages = {791-851},
          doi = {10.1111/j.1365-2966.2009.15715.x},
archivePrefix = {arXiv},
       eprint = {0901.4107},
 primaryClass = {astro-ph.CO},
       adsurl = {https://ui.adsabs.harvard.edu/abs/2010MNRAS.401..791S}
}

@ARTICLE{Pillepich+2019MNRAS.490.3196P,
       author = {{Pillepich}, Annalisa and {Nelson}, Dylan and {Springel}, Volker and {Pakmor}, R{\"u}diger and {Torrey}, Paul and {Weinberger}, Rainer and {Vogelsberger}, Mark and {Marinacci}, Federico and {Genel}, Shy and {van der Wel}, Arjen and {Hernquist}, Lars},
        title = "{First results from the TNG50 simulation: the evolution of stellar and gaseous discs across cosmic time}",
      journal = {\mnras},
         year = 2019,
        month = dec,
       volume = {490},
       number = {3},
        pages = {3196-3233},
          doi = {10.1093/mnras/stz2338},
archivePrefix = {arXiv},
       eprint = {1902.05553},
 primaryClass = {astro-ph.GA},
       adsurl = {https://ui.adsabs.harvard.edu/abs/2019MNRAS.490.3196P}
}

@ARTICLE{Nelson+2019MNRAS.490.3234N,
       author = {{Nelson}, Dylan and {Pillepich}, Annalisa and {Springel}, Volker and {Pakmor}, R{\"u}diger and {Weinberger}, Rainer and {Genel}, Shy and {Torrey}, Paul and {Vogelsberger}, Mark and {Marinacci}, Federico and {Hernquist}, Lars},
        title = "{First results from the TNG50 simulation: galactic outflows driven by supernovae and black hole feedback}",
      journal = {\mnras},
         year = 2019,
        month = dec,
       volume = {490},
       number = {3},
        pages = {3234-3261},
          doi = {10.1093/mnras/stz2306},
archivePrefix = {arXiv},
       eprint = {1902.05554},
 primaryClass = {astro-ph.GA},
       adsurl = {https://ui.adsabs.harvard.edu/abs/2019MNRAS.490.3234N}
}

@ARTICLE{Sales+2020MNRAS.494.1848S,
       author = {{Sales}, Laura V. and {Navarro}, Julio F. and {Pe{\~n}afiel}, Louis and {Peng}, Eric W. and {Lim}, Sungsoon and {Hernquist}, Lars},
        title = "{The formation of ultradiffuse galaxies in clusters}",
      journal = {\mnras},
         year = 2020,
        month = may,
       volume = {494},
       number = {2},
        pages = {1848-1858},
          doi = {10.1093/mnras/staa854},
archivePrefix = {arXiv},
       eprint = {1909.01347},
 primaryClass = {astro-ph.CO},
       adsurl = {https://ui.adsabs.harvard.edu/abs/2020MNRAS.494.1848S}
}

@ARTICLE{Wang+2023Natur.623..296W,
       author = {{Wang}, Kaixiang and {Peng}, Eric W. and {Liu}, Chengze and {Mihos}, J. Christopher and {C{\^o}t{\'e}}, Patrick and {Ferrarese}, Laura and {Taylor}, Matthew A. and {Blakeslee}, John P. and {Cuillandre}, Jean-Charles and {Duc}, Pierre-Alain and {Guhathakurta}, Puragra and {Gwyn}, Stephen and {Ko}, Youkyung and {Lan{\c{c}}on}, Ariane and {Lim}, Sungsoon and {MacArthur}, Lauren A. and {Puzia}, Thomas and {Roediger}, Joel and {Sales}, Laura V. and {S{\'a}nchez-Janssen}, Rub{\'e}n and {Spengler}, Chelsea and {Toloba}, Elisa and {Zhang}, Hongxin and {Zhu}, Mingcheng},
        title = "{An evolutionary continuum from nucleated dwarf galaxies to star clusters}",
      journal = {\nat},
         year = 2023,
        month = nov,
       volume = {623},
       number = {7986},
        pages = {296-300},
          doi = {10.1038/s41586-023-06650-z},
archivePrefix = {arXiv},
       eprint = {2311.05448},
 primaryClass = {astro-ph.GA},
       adsurl = {https://ui.adsabs.harvard.edu/abs/2023Natur.623..296W}
}

@ARTICLE{Norris+2014MNRAS.443.1151N,
       author = {{Norris}, Mark A. and {Kannappan}, Sheila J. and {Forbes}, Duncan A. and {Romanowsky}, Aaron J. and {Brodie}, Jean P. and {Faifer}, Favio Ra{\'u}l and {Huxor}, Avon and {Maraston}, Claudia and {Moffett}, Amanda J. and {Penny}, Samantha J. and {Pota}, Vincenzo and {Smith-Castelli}, Anal{\'\i}a and {Strader}, Jay and {Bradley}, David and {Eckert}, Kathleen D. and {Fohring}, Dora and {McBride}, JoEllen and {Stark}, David V. and {Vaduvescu}, Ovidiu},
        title = "{The AIMSS Project - I. Bridging the star cluster-galaxy divide$^{★}${\textdagger}{\textdaggerdbl}{\textsection}{\textparagraph}}",
      journal = {\mnras},
         year = 2014,
        month = sep,
       volume = {443},
       number = {2},
        pages = {1151-1172},
          doi = {10.1093/mnras/stu1186},
archivePrefix = {arXiv},
       eprint = {1406.6065},
 primaryClass = {astro-ph.GA},
       adsurl = {https://ui.adsabs.harvard.edu/abs/2014MNRAS.443.1151N}
}

@ARTICLE{Trujillo+2006ApJ...650...18T,
       author = {{Trujillo}, Ignacio and {F{\"o}rster Schreiber}, Natascha M. and {Rudnick}, Gregory and {Barden}, Marco and {Franx}, Marijn and {Rix}, Hans-Walter and {Caldwell}, J.~A.~R. and {McIntosh}, Daniel H. and {Toft}, Sune and {H{\"a}ussler}, Boris and {Zirm}, Andrew and {van Dokkum}, Pieter G. and {Labb{\'e}}, Ivo and {Moorwood}, Alan and {R{\"o}ttgering}, Huub and {van der Wel}, Arjen and {van der Werf}, Paul and {van Starkenburg}, Lottie},
        title = "{The Size Evolution of Galaxies since z\raisebox{-0.5ex}\textasciitilde3: Combining SDSS, GEMS, and FIRES}",
      journal = {\apj},
         year = 2006,
        month = oct,
       volume = {650},
       number = {1},
        pages = {18-41},
          doi = {10.1086/506464},
archivePrefix = {arXiv},
       eprint = {astro-ph/0504225},
 primaryClass = {astro-ph},
       adsurl = {https://ui.adsabs.harvard.edu/abs/2006ApJ...650...18T}
}

@ARTICLE{VanDerWel+2014ApJ...788...28V,
       author = {{van der Wel}, A. and {Franx}, M. and {van Dokkum}, P.~G. and {Skelton}, R.~E. and {Momcheva}, I.~G. and {Whitaker}, K.~E. and {Brammer}, G.~B. and {Bell}, E.~F. and {Rix}, H.-W. and {Wuyts}, S. and {Ferguson}, H.~C. and {Holden}, B.~P. and {Barro}, G. and {Koekemoer}, A.~M. and {Chang}, Yu-Yen and {McGrath}, E.~J. and {H{\"a}ussler}, B. and {Dekel}, A. and {Behroozi}, P. and {Fumagalli}, M. and {Leja}, J. and {Lundgren}, B.~F. and {Maseda}, M.~V. and {Nelson}, E.~J. and {Wake}, D.~A. and {Patel}, S.~G. and {Labb{\'e}}, I. and {Faber}, S.~M. and {Grogin}, N.~A. and {Kocevski}, D.~D.},
        title = "{3D-HST+CANDELS: The Evolution of the Galaxy Size-Mass Distribution since z = 3}",
      journal = {\apj},
         year = 2014,
        month = jun,
       volume = {788},
       number = {1},
          eid = {28},
        pages = {28},
          doi = {10.1088/0004-637X/788/1/28},
archivePrefix = {arXiv},
       eprint = {1404.2844},
 primaryClass = {astro-ph.GA},
       adsurl = {https://ui.adsabs.harvard.edu/abs/2014ApJ...788...28V}
}

@ARTICLE{Shibuya+2015ApJS..219...15S,
       author = {{Shibuya}, Takatoshi and {Ouchi}, Masami and {Harikane}, Yuichi},
        title = "{Morphologies of {\ensuremath{\sim}}190,000 Galaxies at z = 0-10 Revealed with HST Legacy Data. I. Size Evolution}",
      journal = {\apjs},
         year = 2015,
        month = aug,
       volume = {219},
       number = {2},
          eid = {15},
        pages = {15},
          doi = {10.1088/0067-0049/219/2/15},
archivePrefix = {arXiv},
       eprint = {1503.07481},
 primaryClass = {astro-ph.GA},
       adsurl = {https://ui.adsabs.harvard.edu/abs/2015ApJS..219...15S}
}

@ARTICLE{Genel+2018MNRAS.474.3976G,
       author = {{Genel}, Shy and {Nelson}, Dylan and {Pillepich}, Annalisa and {Springel}, Volker and {Pakmor}, R{\"u}diger and {Weinberger}, Rainer and {Hernquist}, Lars and {Naiman}, Jill and {Vogelsberger}, Mark and {Marinacci}, Federico and {Torrey}, Paul},
        title = "{The size evolution of star-forming and quenched galaxies in the IllustrisTNG simulation}",
      journal = {\mnras},
         year = 2018,
        month = mar,
       volume = {474},
       number = {3},
        pages = {3976-3996},
          doi = {10.1093/mnras/stx3078},
archivePrefix = {arXiv},
       eprint = {1707.05327},
 primaryClass = {astro-ph.GA},
       adsurl = {https://ui.adsabs.harvard.edu/abs/2018MNRAS.474.3976G}
}

@ARTICLE{vanDokkum+2015ApJ...798L..45V,
       author = {{van Dokkum}, Pieter G. and {Abraham}, Roberto and {Merritt}, Allison and {Zhang}, Jielai and {Geha}, Marla and {Conroy}, Charlie},
        title = "{Forty-seven Milky Way-sized, Extremely Diffuse Galaxies in the Coma Cluster}",
      journal = {\apjl},
         year = 2015,
        month = jan,
       volume = {798},
       number = {2},
          eid = {L45},
        pages = {L45},
          doi = {10.1088/2041-8205/798/2/L45},
archivePrefix = {arXiv},
       eprint = {1410.8141},
 primaryClass = {astro-ph.GA},
       adsurl = {https://ui.adsabs.harvard.edu/abs/2015ApJ...798L..45V}
}

@ARTICLE{Sandage+Binggeli1984AJ.....89..919S,
       author = {{Sandage}, A. and {Binggeli}, B.},
        title = "{Studies of the Virgo cluster. III. A classification system and an illustrated Atlas of Virgo cluster dwarf galaxies.}",
      journal = {\aj},
         year = 1984,
        month = jul,
       volume = {89},
        pages = {919-931},
          doi = {10.1086/113588},
       adsurl = {https://ui.adsabs.harvard.edu/abs/1984AJ.....89..919S}
}

@ARTICLE{Conselice+2003AJ....125...66C,
       author = {{Conselice}, Christopher J. and {Gallagher}, John S., III and {Wyse}, Rosemary F.~G.},
        title = "{Galaxy Populations and Evolution in Clusters. III. The Origin of Low-Mass Galaxies in Clusters: Constraints from Stellar Populations}",
      journal = {\aj},
         year = 2003,
        month = jan,
       volume = {125},
       number = {1},
        pages = {66-85},
          doi = {10.1086/345385},
archivePrefix = {arXiv},
       eprint = {astro-ph/0210080},
 primaryClass = {astro-ph},
       adsurl = {https://ui.adsabs.harvard.edu/abs/2003AJ....125...66C}
}

@ARTICLE{Impey+1988ApJ...330..634I,
       author = {{Impey}, Chris and {Bothun}, Greg and {Malin}, David},
        title = "{Virgo Dwarfs: New Light on Faint Galaxies}",
      journal = {\apj},
         year = 1988,
        month = jul,
       volume = {330},
        pages = {634},
          doi = {10.1086/166500},
       adsurl = {https://ui.adsabs.harvard.edu/abs/1988ApJ...330..634I}
}

@ARTICLE{Koda+2015ApJ...807L...2K,
       author = {{Koda}, Jin and {Yagi}, Masafumi and {Yamanoi}, Hitomi and {Komiyama}, Yutaka},
        title = "{Approximately a Thousand Ultra-diffuse Galaxies in the Coma Cluster}",
      journal = {\apjl},
         year = 2015,
        month = jul,
       volume = {807},
       number = {1},
          eid = {L2},
        pages = {L2},
          doi = {10.1088/2041-8205/807/1/L2},
archivePrefix = {arXiv},
       eprint = {1506.01712},
 primaryClass = {astro-ph.GA},
       adsurl = {https://ui.adsabs.harvard.edu/abs/2015ApJ...807L...2K}
}

@ARTICLE{VanDerBurg+2017A&A...607A..79V,
       author = {{van der Burg}, Remco F.~J. and {Hoekstra}, Henk and {Muzzin}, Adam and {Sif{\'o}n}, Crist{\'o}bal and {Viola}, Massimo and {Bremer}, Malcolm N. and {Brough}, Sarah and {Driver}, Simon P. and {Erben}, Thomas and {Heymans}, Catherine and {Hildebrandt}, Hendrik and {Holwerda}, Benne W. and {Klaes}, Dominik and {Kuijken}, Konrad and {McGee}, Sean and {Nakajima}, Reiko and {Napolitano}, Nicola and {Norberg}, Peder and {Taylor}, Edward N. and {Valentijn}, Edwin},
        title = "{The abundance of ultra-diffuse galaxies from groups to clusters. UDGs are relatively more common in more massive haloes}",
      journal = {\aap},
         year = 2017,
        month = nov,
       volume = {607},
          eid = {A79},
        pages = {A79},
          doi = {10.1051/0004-6361/201731335},
archivePrefix = {arXiv},
       eprint = {1706.02704},
 primaryClass = {astro-ph.GA},
       adsurl = {https://ui.adsabs.harvard.edu/abs/2017A&A...607A..79V}
}

@ARTICLE{Bautista+2023ApJS..267...10B,
       author = {{Bautista}, Jose Miguel G. and {Koda}, Jin and {Yagi}, Masafumi and {Komiyama}, Yutaka and {Yamanoi}, Hitomi},
        title = "{Ultradiffuse Galaxies (UDGs) with Hyper Suprime-Cam. I. Revised Catalog of Coma Cluster UDGs}",
      journal = {\apjs},
         year = 2023,
        month = jul,
       volume = {267},
       number = {1},
          eid = {10},
        pages = {10},
          doi = {10.3847/1538-4365/acd3e7},
archivePrefix = {arXiv},
       eprint = {2307.07141},
 primaryClass = {astro-ph.GA},
       adsurl = {https://ui.adsabs.harvard.edu/abs/2023ApJS..267...10B}
}

@ARTICLE{Yagi+2016ApJS..225...11Y,
       author = {{Yagi}, Masafumi and {Koda}, Jin and {Komiyama}, Yutaka and {Yamanoi}, Hitomo},
        title = "{Catalog of Ultra-diffuse Galaxies in the Coma Clusters from Subaru Imaging Data}",
      journal = {\apjs},
         year = 2016,
        month = jul,
       volume = {225},
       number = {1},
          eid = {11},
        pages = {11},
          doi = {10.3847/0067-0049/225/1/11},
       adsurl = {https://ui.adsabs.harvard.edu/abs/2016ApJS..225...11Y}
}

@ARTICLE{Gannon+2024MNRAS.531.1856G,
       author = {{Gannon}, Jonah S. and {Ferr{\'e}-Mateu}, Anna and {Forbes}, Duncan A. and {Brodie}, Jean P. and {Buzzo}, Maria Luisa and {Romanowsky}, Aaron J.},
        title = "{A Catalogue and analysis of ultra-diffuse galaxy spectroscopic properties}",
      journal = {\mnras},
         year = 2024,
        month = jun,
       volume = {531},
       number = {1},
        pages = {1856-1869},
          doi = {10.1093/mnras/stae1287},
archivePrefix = {arXiv},
       eprint = {2405.09104},
 primaryClass = {astro-ph.GA},
       adsurl = {https://ui.adsabs.harvard.edu/abs/2024MNRAS.531.1856G}
}

@ARTICLE{Zaritsky+2023ApJS..267...27Z,
       author = {{Zaritsky}, Dennis and {Donnerstein}, Richard and {Dey}, Arjun and {Karunakaran}, Ananthan and {Kadowaki}, Jennifer and {Khim}, Donghyeon J. and {Spekkens}, Kristine and {Zhang}, Huanian},
        title = "{Systematically Measuring Ultra-diffuse Galaxies (SMUDGes). V. The Complete SMUDGes Catalog and the Nature of Ultradiffuse Galaxies}",
      journal = {\apjs},
         year = 2023,
        month = aug,
       volume = {267},
       number = {2},
          eid = {27},
        pages = {27},
          doi = {10.3847/1538-4365/acdd71},
archivePrefix = {arXiv},
       eprint = {2306.01524},
 primaryClass = {astro-ph.GA},
       adsurl = {https://ui.adsabs.harvard.edu/abs/2023ApJS..267...27Z}
}

@ARTICLE{Zaritsky+2022ApJS..261...11Z,
       author = {{Zaritsky}, Dennis and {Donnerstein}, Richard and {Karunakaran}, Ananthan and {Barbosa}, C.~E. and {Dey}, Arjun and {Kadowaki}, Jennifer and {Spekkens}, Kristine and {Zhang}, Huanian},
        title = "{Systematically Measuring Ultra-diffuse Galaxies (SMUDGes). III. The Southern SMUDGes Catalog}",
      journal = {\apjs},
         year = 2022,
        month = aug,
       volume = {261},
       number = {2},
          eid = {11},
        pages = {11},
          doi = {10.3847/1538-4365/ac6ceb},
archivePrefix = {arXiv},
       eprint = {2205.02193},
 primaryClass = {astro-ph.GA},
       adsurl = {https://ui.adsabs.harvard.edu/abs/2022ApJS..261...11Z}
}

@article{Gannon+2026, 
            title={The Dawes review 14: A decade of ultra-diffuse galaxies},
            volume={43},
            doi={10.1017/pasa.2026.10169},
            journal={\pasa}, 
            author={Gannon, Jonah S. and Ferré-Mateu, Anna and Forbes, Duncan A},
            year={2026},
            pages={e031}}

@ARTICLE{DiCintio+2017MNRAS.466L...1D,
       author = {{Di Cintio}, Arianna and {Brook}, Chris B. and {Dutton}, Aaron A. and {Macci{\`o}}, Andrea V. and {Obreja}, Aura and {Dekel}, Avishai},
        title = "{NIHAO - XI. Formation of ultra-diffuse galaxies by outflows}",
      journal = {\mnras},
         year = 2017,
        month = mar,
       volume = {466},
       number = {1},
        pages = {L1-L6},
          doi = {10.1093/mnrasl/slw210},
archivePrefix = {arXiv},
       eprint = {1608.01327},
 primaryClass = {astro-ph.GA},
       adsurl = {https://ui.adsabs.harvard.edu/abs/2017MNRAS.466L...1D}
}

@ARTICLE{Chan+2018MNRAS.478..906C,
       author = {{Chan}, T.~K. and {Kere{\v{s}}}, D. and {Wetzel}, A. and {Hopkins}, P.~F. and {Faucher-Gigu{\`e}re}, C. -A. and {El-Badry}, K. and {Garrison-Kimmel}, S. and {Boylan-Kolchin}, M.},
        title = "{The origin of ultra diffuse galaxies: stellar feedback and quenching}",
      journal = {\mnras},
         year = 2018,
        month = jul,
       volume = {478},
       number = {1},
        pages = {906-925},
          doi = {10.1093/mnras/sty1153},
archivePrefix = {arXiv},
       eprint = {1711.04788},
 primaryClass = {astro-ph.GA},
       adsurl = {https://ui.adsabs.harvard.edu/abs/2018MNRAS.478..906C}
}

@article{Jiang+2019,
    author = {Jiang, Fangzhou and Dekel, Avishai and Freundlich, Jonathan and Romanowsky, Aaron J and Dutton, Aaron A and Macciò, Andrea V and Di Cintio, Arianna},
    title = "{Formation of ultra-diffuse galaxies in the field and in galaxy groups}",
    journal = {\mnras},
    volume = {487},
    number = {4},
    pages = {5272-5290},
    year = {2019},
    month = {06},
    issn = {0035-8711},
    doi = {10.1093/mnras/stz1499},
    url = {https://doi.org/10.1093/mnras/stz1499},
    eprint = {https://academic.oup.com/mnras/article-pdf/487/4/5272/28893626/stz1499.pdf},
}

@ARTICLE{Jackson+2021MNRAS.502.4262J,
       author = {{Jackson}, R.~A. and {Martin}, G. and {Kaviraj}, S. and {Rams{\o}y}, M. and {Devriendt}, J.~E.~G. and {Sedgwick}, T. and {Laigle}, C. and {Choi}, H. and {Beckmann}, R.~S. and {Volonteri}, M. and {Dubois}, Y. and {Pichon}, C. and {Yi}, S.~K. and {Slyz}, A. and {Kraljic}, K. and {Kimm}, T. and {Peirani}, S. and {Baldry}, I.},
        title = "{The origin of low-surface-brightness galaxies in the dwarf regime}",
      journal = {\mnras},
         year = 2021,
        month = apr,
       volume = {502},
       number = {3},
        pages = {4262-4276},
          doi = {10.1093/mnras/stab077},
archivePrefix = {arXiv},
       eprint = {2007.06581},
 primaryClass = {astro-ph.GA},
       adsurl = {https://ui.adsabs.harvard.edu/abs/2021MNRAS.502.4262J}
}

@ARTICLE{Tremmel+2019MNRAS.483.3336T,
       author = {{Tremmel}, M. and {Quinn}, T.~R. and {Ricarte}, A. and {Babul}, A. and {Chadayammuri}, U. and {Natarajan}, P. and {Nagai}, D. and {Pontzen}, A. and {Volonteri}, M.},
        title = "{Introducing ROMULUSC: a cosmological simulation of a galaxy cluster with an unprecedented resolution}",
      journal = {\mnras},
         year = 2019,
        month = mar,
       volume = {483},
       number = {3},
        pages = {3336-3362},
          doi = {10.1093/mnras/sty3336},
archivePrefix = {arXiv},
       eprint = {1806.01282},
 primaryClass = {astro-ph.GA},
       adsurl = {https://ui.adsabs.harvard.edu/abs/2019MNRAS.483.3336T}
}

@ARTICLE{Tremmel+2020MNRAS.497.2786T,
       author = {{Tremmel}, M. and {Wright}, A.~C. and {Brooks}, A.~M. and {Munshi}, F. and {Nagai}, D. and {Quinn}, T.~R.},
        title = "{The formation of ultradiffuse galaxies in the RomulusC galaxy cluster simulation}",
      journal = {\mnras},
         year = 2020,
        month = sep,
       volume = {497},
       number = {3},
        pages = {2786-2810},
          doi = {10.1093/mnras/staa2015},
archivePrefix = {arXiv},
       eprint = {1908.05684},
 primaryClass = {astro-ph.GA},
       adsurl = {https://ui.adsabs.harvard.edu/abs/2020MNRAS.497.2786T}
}

@ARTICLE{Hilker+1999A&AS..134...75H,
       author = {{Hilker}, M. and {Infante}, L. and {Vieira}, G. and {Kissler-Patig}, M. and {Richtler}, T.},
        title = "{The central region of the Fornax cluster. II. Spectroscopy and radial velocities of member and background galaxies}",
      journal = {\aaps},
         year = 1999,
        month = jan,
       volume = {134},
        pages = {75-86},
          doi = {10.1051/aas:1999434},
archivePrefix = {arXiv},
       eprint = {astro-ph/9807144},
 primaryClass = {astro-ph},
       adsurl = {https://ui.adsabs.harvard.edu/abs/1999A&AS..134...75H}
}

@ARTICLE{Drinkwater+2000PASA...17..227D,
       author = {{Drinkwater}, M.~J. and {Jones}, J.~B. and {Gregg}, M.~D. and {Phillipps}, S.},
        title = "{Compact Stellar Systems in the Fornax Cluster: Super-massive Star Clusters or Extremely Compact Dwarf Galaxies?}",
      journal = {\pasa},
         year = 2000,
        month = dec,
       volume = {17},
       number = {3},
        pages = {227-233},
          doi = {10.1071/AS00034},
archivePrefix = {arXiv},
       eprint = {astro-ph/0002003},
 primaryClass = {astro-ph},
       adsurl = {https://ui.adsabs.harvard.edu/abs/2000PASA...17..227D}
}

@ARTICLE{Mieske+2008A&A...487..921M,
       author = {{Mieske}, S. and {Hilker}, M. and {Jord{\'a}n}, A. and {Infante}, L. and {Kissler-Patig}, M. and {Rejkuba}, M. and {Richtler}, T. and {C{\^o}t{\'e}}, P. and {Baumgardt}, H. and {West}, M.~J. and {Ferrarese}, L. and {Peng}, E.~W.},
        title = "{The nature of UCDs: Internal dynamics from an expanded sample and homogeneous database}",
      journal = {\aap},
         year = 2008,
        month = sep,
       volume = {487},
       number = {3},
        pages = {921-935},
          doi = {10.1051/0004-6361:200810077},
archivePrefix = {arXiv},
       eprint = {0806.0374},
 primaryClass = {astro-ph},
       adsurl = {https://ui.adsabs.harvard.edu/abs/2008A&A...487..921M}
}

@ARTICLE{Brodie+2011AJ....142..199B,
       author = {{Brodie}, Jean P. and {Romanowsky}, Aaron J. and {Strader}, Jay and {Forbes}, Duncan A.},
        title = "{The Relationships among Compact Stellar Systems: A Fresh View of Ultracompact Dwarfs}",
      journal = {\aj},
         year = 2011,
        month = dec,
       volume = {142},
       number = {6},
          eid = {199},
        pages = {199},
          doi = {10.1088/0004-6256/142/6/199},
archivePrefix = {arXiv},
       eprint = {1109.5696},
 primaryClass = {astro-ph.CO},
       adsurl = {https://ui.adsabs.harvard.edu/abs/2011AJ....142..199B}
}

@ARTICLE{Norris+2011MNRAS.414..739N,
       author = {{Norris}, Mark A. and {Kannappan}, Sheila J.},
        title = "{The ubiquity and dual nature of ultra-compact dwarfs}",
      journal = {\mnras},
         year = 2011,
        month = jun,
       volume = {414},
       number = {1},
        pages = {739-758},
          doi = {10.1111/j.1365-2966.2011.18440.x},
archivePrefix = {arXiv},
       eprint = {1102.0001},
 primaryClass = {astro-ph.CO},
       adsurl = {https://ui.adsabs.harvard.edu/abs/2011MNRAS.414..739N}
}

@ARTICLE{Fellhauer+Kroupa2002MNRAS.330..642F,
       author = {{Fellhauer}, Michael and {Kroupa}, Pavel},
        title = "{The formation of ultracompact dwarf galaxies}",
      journal = {\mnras},
         year = 2002,
        month = mar,
       volume = {330},
       number = {3},
        pages = {642-650},
          doi = {10.1046/j.1365-8711.2002.05087.x},
archivePrefix = {arXiv},
       eprint = {astro-ph/0110621},
 primaryClass = {astro-ph},
       adsurl = {https://ui.adsabs.harvard.edu/abs/2002MNRAS.330..642F}
}

@ARTICLE{bekki2001,
       author = {{Bekki}, Kenji and {Couch}, Warrick J. and {Drinkwater}, Michael J. and {Gregg}, Michael D.},
        title = "{A New Formation Model for M32: A Threshed Early-Type Spiral Galaxy?}",
      journal = {\apjl},
         year = 2001,
        month = aug,
       volume = {557},
       number = {1},
        pages = {L39-L42},
          doi = {10.1086/323075},
archivePrefix = {arXiv},
       eprint = {astro-ph/0107117},
 primaryClass = {astro-ph},
       adsurl = {https://ui.adsabs.harvard.edu/abs/2001ApJ...557L..39B}
}

\begin{appendix}

\section{Candidates}\label{sec:candidates}

\begin{figure}[ht!]
    \centering
    \newcommand{\subfigwidth}{0.48\textwidth}

    \begin{subfigure}{\subfigwidth}
        \includegraphics[width=0.97\columnwidth]{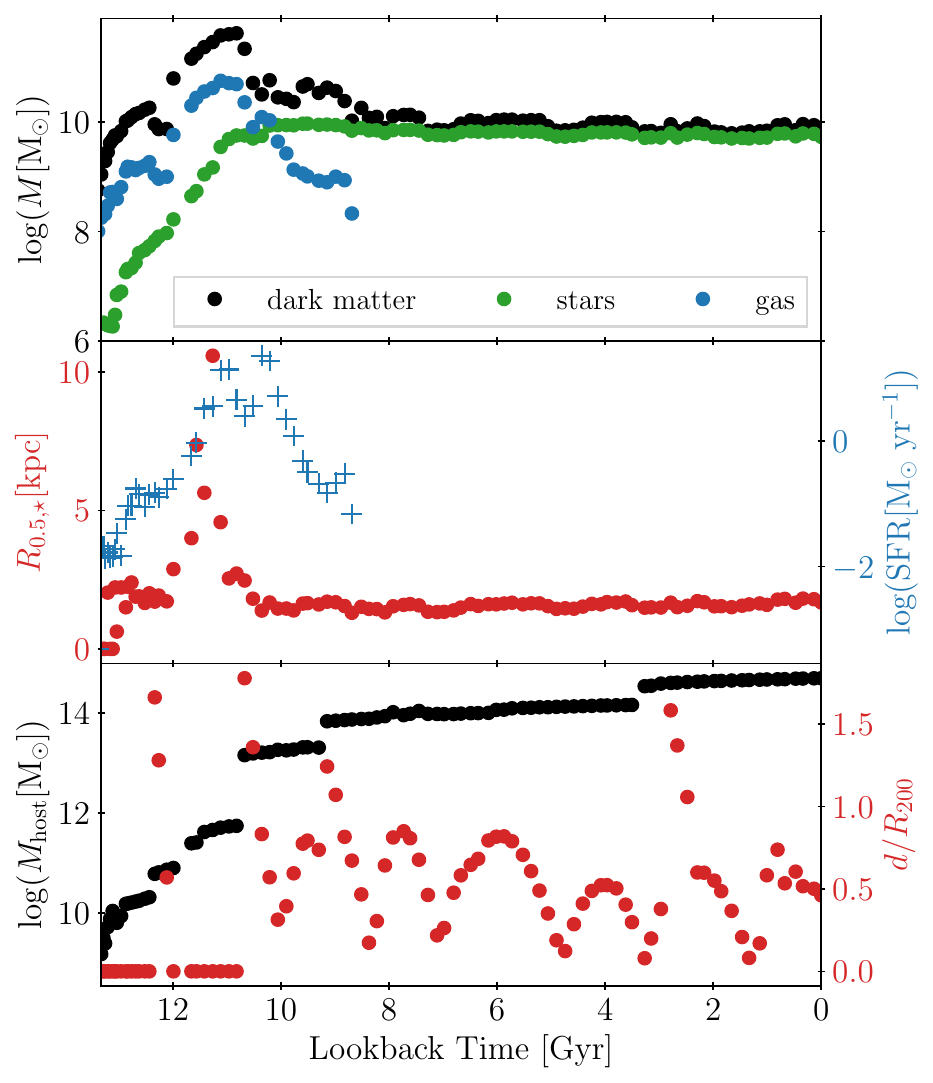}
        \caption{ID17358}
        \label{fig:17358}
    \end{subfigure}
    \hfill 
    \begin{subfigure}{\subfigwidth}
        \includegraphics[width=0.97\columnwidth]{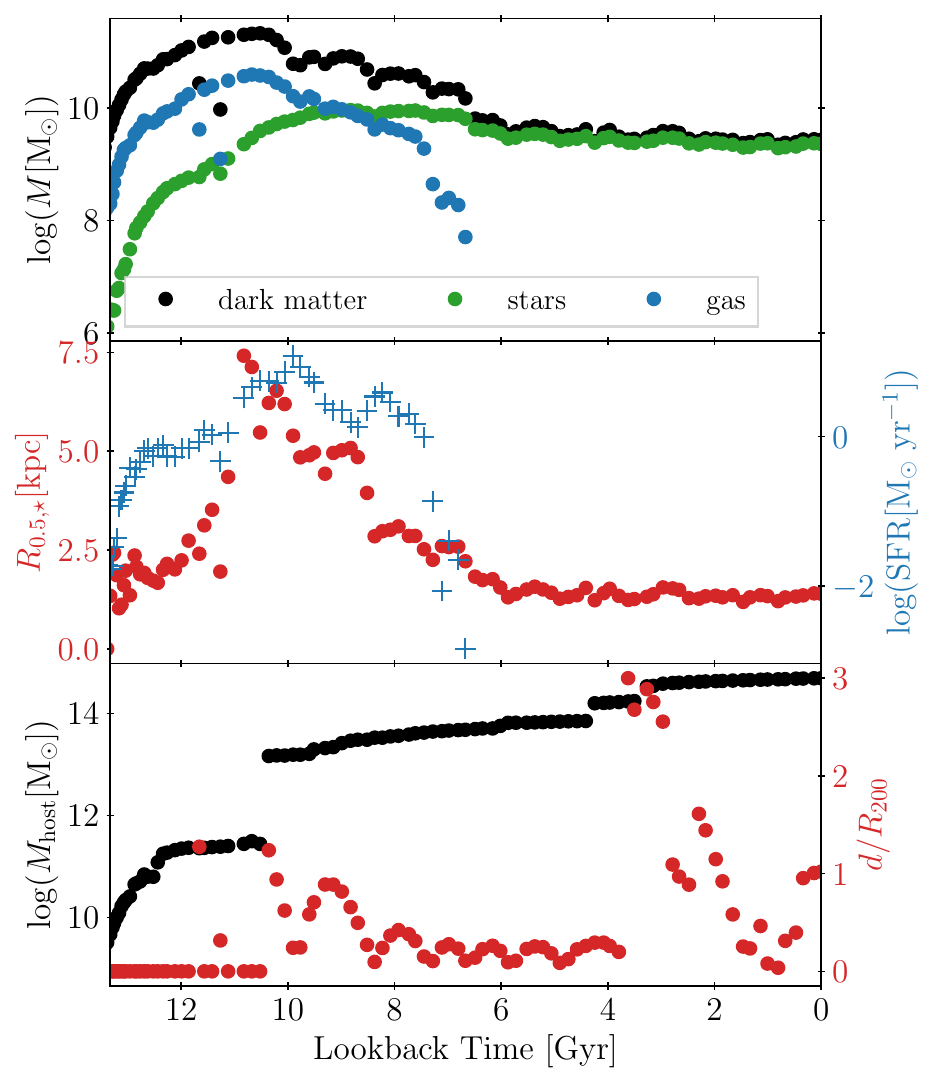}
        \caption{ID17503}
        \label{fig:17503}
    \end{subfigure}

    \caption{Evolutionary histories of CD progenitors that underwent a transient UDG phase. Each subfigure represents a different galaxy, indicated by its present-day \texttt{SubfindID}. Top panels: mass evolution of the individual components of the galaxy (dark matter, stars, and gas). Middle panels: evolution of the stellar half-mass radius (red, left $y$-axis) and SFR (blue, right $y$-axis). Bottom panels: evolution of the host halo mass that represents the mass of the cluster at the present day (black, left $y$-axis), and relative cluster-centric radius (red, right $y$-axis).}
    \label{fig:fourplots+}
\end{figure}

\begin{figure*}[ht!]
    \centering

    \newcommand{\subfigwidth}{0.48\textwidth}    

    \begin{subfigure}{\subfigwidth}
        \includegraphics[width=0.97\columnwidth]{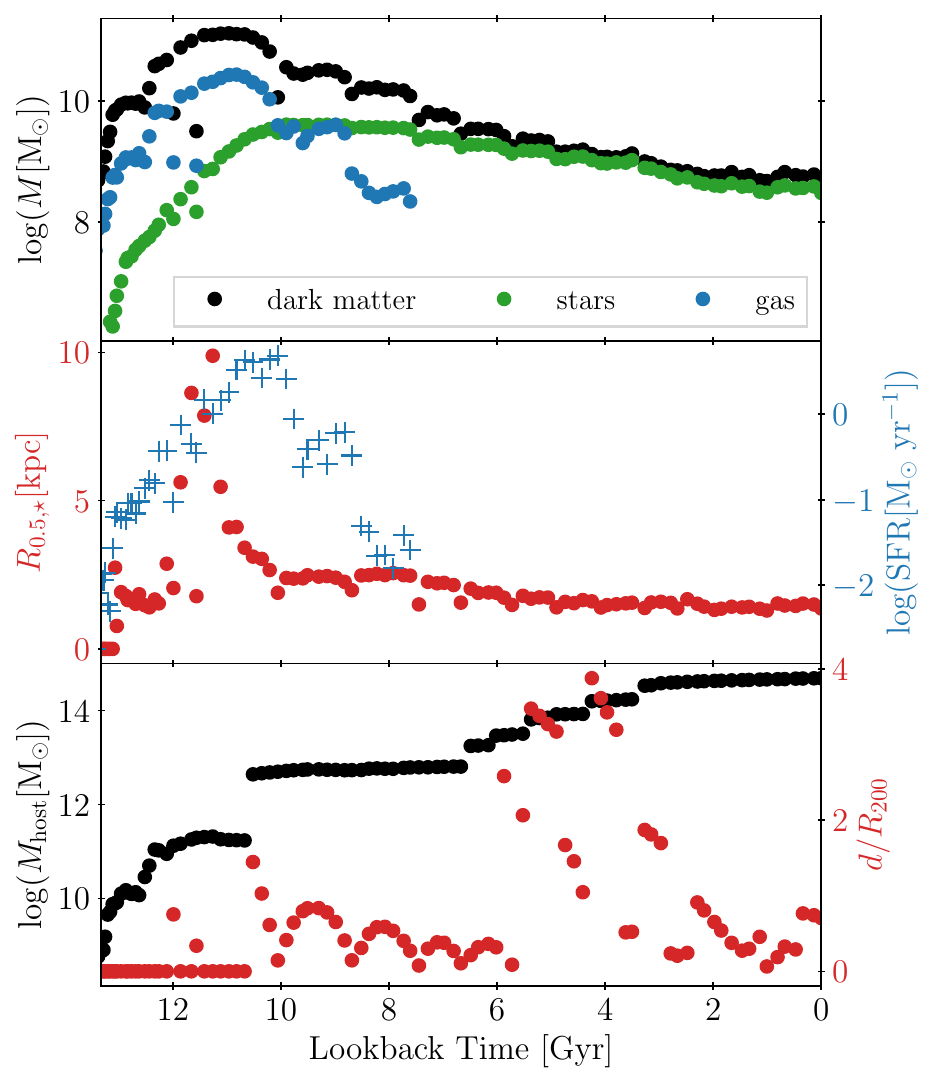}
        \caption{ID18638}
        \label{fig:18638}
    \end{subfigure}
    \hfill
    \begin{subfigure}{\subfigwidth}
        \includegraphics[width=0.97\columnwidth]{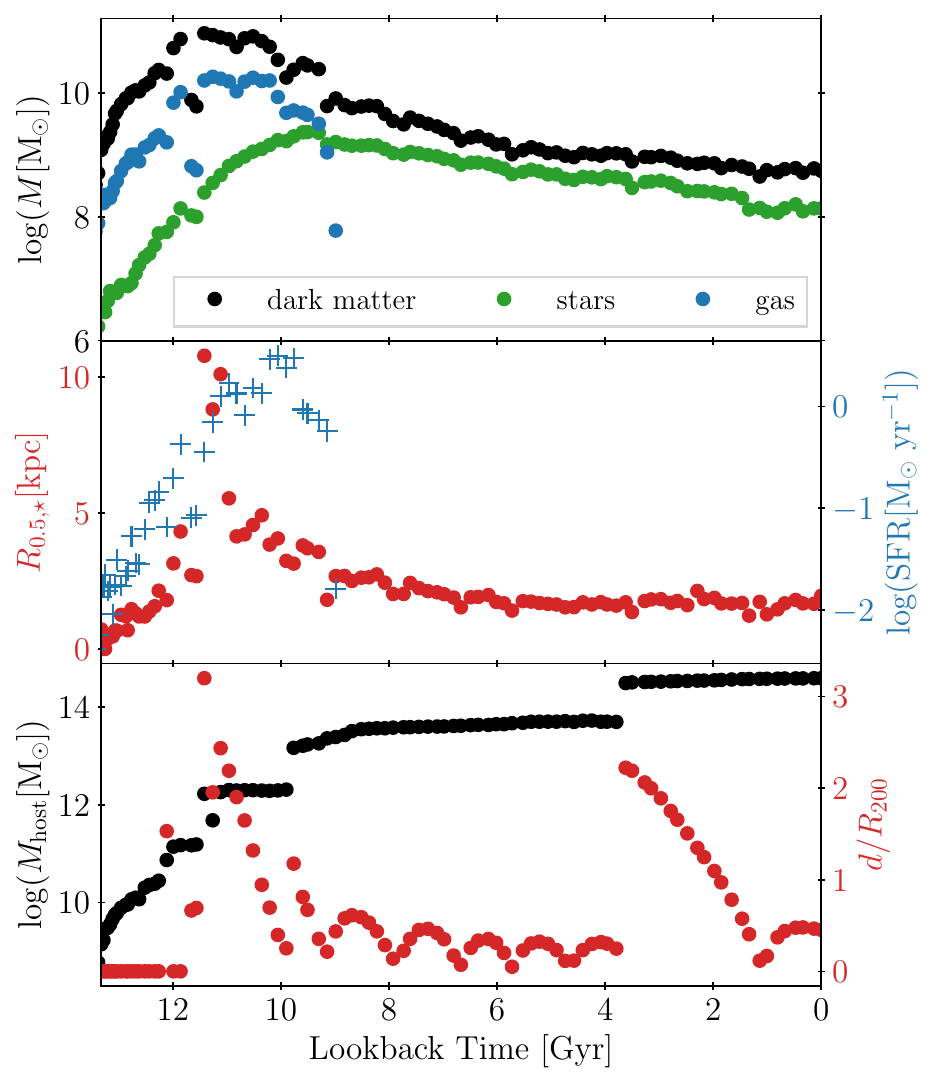}
        \caption{ID33053}
        \label{fig:33053}
    \end{subfigure}

    \vspace{0.5cm}

    \begin{subfigure}{\subfigwidth}
        \includegraphics[width=0.97\columnwidth]{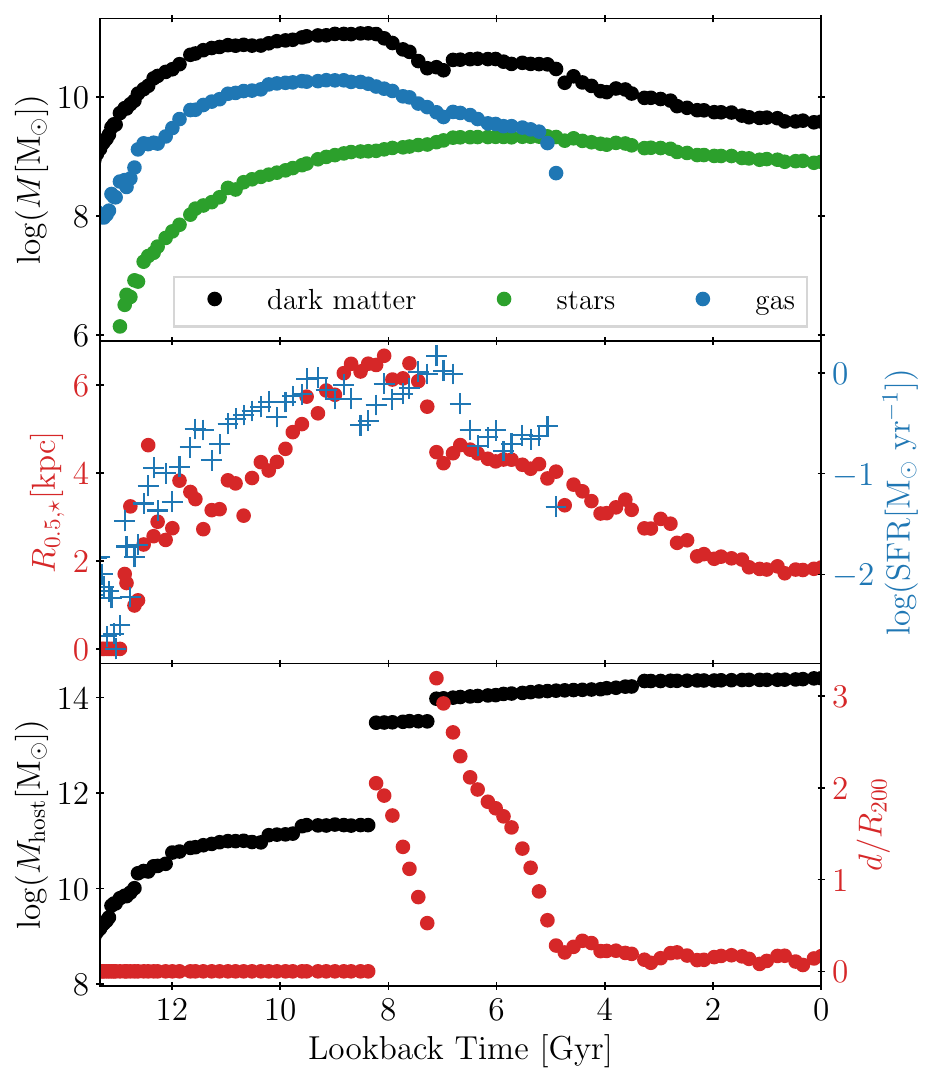}
        \caption{ID88976}
        \label{fig:88976}
    \end{subfigure}
    \hfill
    \begin{subfigure}{\subfigwidth}
        \includegraphics[width=0.97\columnwidth]{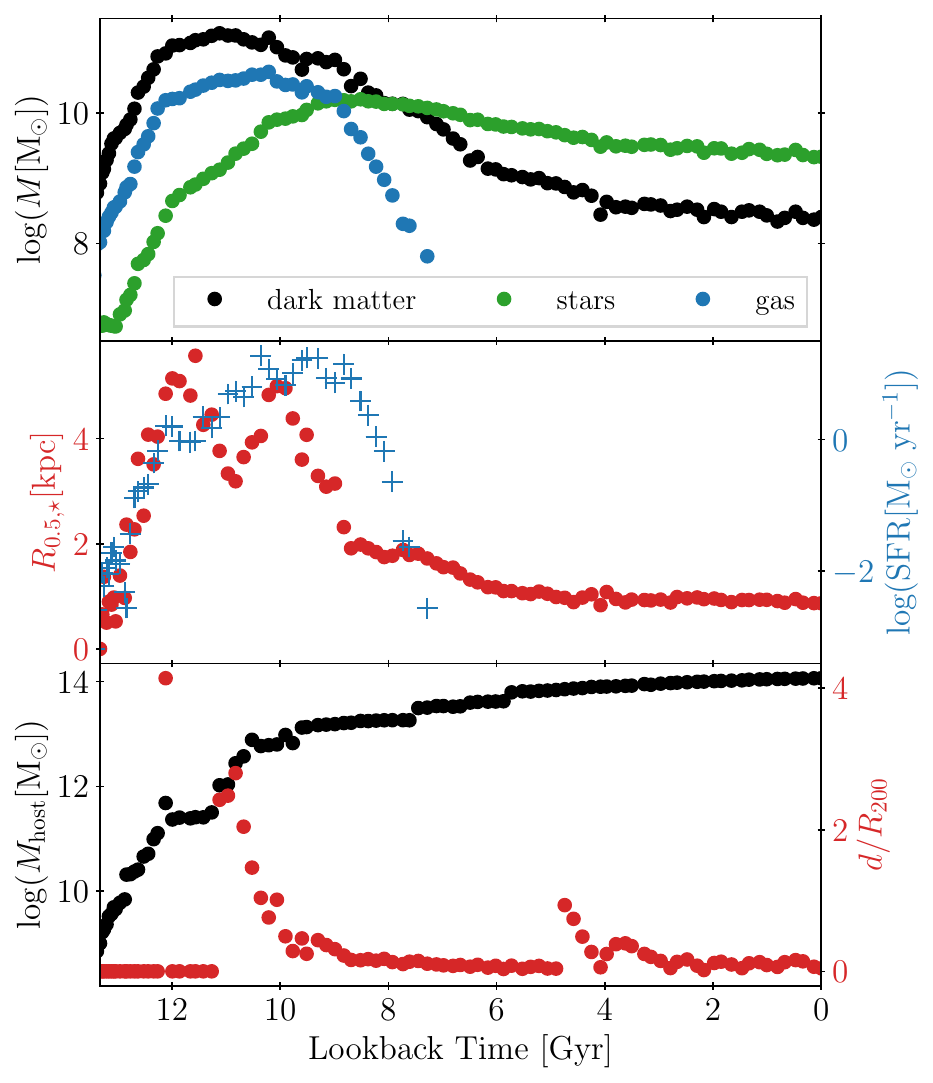}
        \caption{ID131151}
        \label{fig:131151}
    \end{subfigure}

    \caption{Fig.~\ref{fig:fourplots+} Continued.}
    \label{fig:fourplots++}
\end{figure*}

Figure~\ref{fig:fourplots+} and Fig.~\ref{fig:fourplots++} are a continuation of Fig.~\ref{fig:fourplots}, with the same style, arrangement, and notations. 

\end{appendix}

\end{document}